\documentclass[12pt]{iopart}

\usepackage{graphicx}

\usepackage{iopams}  
\usepackage{amssymb}
\usepackage{amsfonts}
\usepackage{setstack} 
\eqnobysec
\usepackage{dsfont} 
\usepackage[dvipsnames]{xcolor}
\usepackage{color}

\newcommand{\be}{\begin{eqnarray}}
\newcommand{\bel}[1]{\begin{equation}\label{#1}}
\newcommand{\ee}{\end{equation}}
\newcommand{\bea}{\begin{eqnarray}}
\newcommand{\eea}{\end{eqnarray}}
\newcommand{\balign}{\begin{align}}
\newcommand{\ealign}{\end{align}}
\newcommand{\ba}{\begin{array}}
\newcommand{\ea}{\end{array}}
\newcommand{\bfig}{\begin{figure}}
\newcommand{\efig}{\end{figure}}

\newcommand{\exval}[1]{\mbox{$\langle \, {#1}\, \rangle$}}

\newcommand{\floor}[1]{\lfloor{#1}\rfloor}

\newcommand{\half}{\frac{1}{2}}

\newcommand{\jrev}{\tilde{\jmath}}

\newcommand{\R}{{\mathbb R}}

\newcommand{\Z}{{\mathbb Z}}
\newcommand{\N}{{\mathbb N}}
 
\newcommand{\T}{{\mathbb T}} 
\newcommand{\I}{{\mathbb I}} 

\newtheorem{theorem}{Theorem}[section]

\newtheorem{definition}[theorem]{Definition}
\newtheorem{proposition}[theorem]{Proposition}

\newtheorem{remark}[theorem]{Remark}

\newcommand{\proof}{\noindent {\it Proof: }}
\def\qed{\hfill$\Box$\par\medskip\par\relax}

\begin{document}

\title{On the phase transition in the sublattice TASEP with 
stochastic blockage}

\author{G M Sch\"utz}

\address{Institut für Biologische Informationsprozesse 5,
Theoretische Physik der Lebenden Materie,
Forschungszentrum J\"ulich, 52425 J\"ulich, Germany}
\ead{g.schuetz@fz-juelich.de}
\vspace{10pt}

\begin{abstract}
We revisit the defect-induced nonequilibrium phase transition from a largely 
homogeneous free-flow phase to a phase-separated congested phase in the 
sublattice totally asymmetric simple exclusion process (TASEP) with 
local deterministic bulk dynamics and a stochastic defect 
that mimicks a random blockage. Exact results are obtained for the 
compressibility and density correlations for a stationary grandcanonical 
ensemble given by the matrix product ansatz. At the critical density the static 
compressibility diverges while in the phase separated state above the critical 
point the compressibility vanishes due to strong non-local correlations. These 
correlations arise from a long range effective interaction between particles that 
appears in the stationary state despite the locality of the microscopic dynamics.
\end{abstract}

\vspace{2pc}

\section{Introduction}
\label{sec:1}

Non-equilibrium phase transitions in one-dimensional driven diffusive systems 
caused by a single static defect bond have a long history of study 
\cite{Wolf90,Jano92,Schu93a,Tang93,Henk94,Sepp01,Baha04} 
and continue to intrigue not only from a statistical physics and probabilistic 
perspective \cite{Schm15,Basu16,Baha18,Szav18,Nejj18} but also because
of their recently recognized significance for biological transport by molecular 
motors \cite{Appe15,Mish16,Ghos19,Jind20}. The general picture is that at a 
critical density $\rho_c$ of driven particles there is a defect-induced 
nonequilibrium phase transition from a spatially homogeneous ``free-flow'' 
phase for $\rho<\rho_c$ to a ``congested phase'' for $\rho>\rho_c$ with two 
coexisting low density and high-density segments, corresponding to the 
formation of a macroscopic ``traffic jam'' upstream of the blockage bond. 

Thus this phenomenon can be regarded as a nonequilibrium analog of phase 
separation \cite{Rama02,Kafr03,Chak16}. In the phase separated state the 
stationary particle current becomes independent of the total conserved particle 
density. Increasing the total density enlarges the size of the high-density 
segment rather than changing the current whose maximally attainable value is 
limited by the blockage strength. The high-density segment and the low-density 
segment are separated by a domain wall which is sharp even on microscopic 
scale and represents a microscopic realization of what on macroscopic scale 
constitutes a shock, i.e., a discontinuity in the macroscopic density profile 
along the system. Such a stable domain wall is generally believed to perform 
a random walk, see e.g. analytical results for the continuous-time asymmetric 
simple exclusion process obtained by a variety of different methods 
\cite{Ferr94,Dudz00,Beli02,Kreb03,deGi06,Bala10,Beli18}.

Due to particle number conservation, the shock position in a finite system is 
confined to a region compatible with the conserved total density. Moreover, 
long-range correlations between the upstream and downstream regions to the 
left and right of the blockage respectively were postulated to explain 
numerically observed fluctuations of the shock position around its stationary 
mean \cite{Jano92}. To elucidate the phase transition and associated long 
range correlations further, we consider the sublattice totally asymmetric 
simple exclusion process (dsTASEP) with deterministic bulk dynamics and local 
dynamical randomness introduced by a defect \cite{Schu93a}, informally 
defined for the one-dimensional lattice with $L$ sites as follows.

We use the convention to denote by $\N$ the set of positive integers and by 
$\N_0$ the set of nonnegative integers. The set $\T_L = \{1,\dots ,L\}$ 
refers to the lattice. The state $\eta=(\eta_1,\dots,\eta_L)$ of the dsTASEP 
is represented at any given time by the local occupation numbers 
$\eta_k \in \{0,1\}$. For $\eta_k = 1$, we say that site $k\in\T_L$ is 
occupied by a particle, thus encoding hard-core repulsion that forbids double 
occupancy of a site. For $\eta_k = 0$ we say that site $k$ is empty or, 
alternatively, occupied by a hole. Correspondingly, 
\begin{equation}
\bar{\eta}_k := 1 - \eta_k 
\end{equation}
are the hole occupation numbers.

The dsTASEP is a stochastic cellular automaton evolving in discrete time steps
$t \in \N_0$ and is described by the sequence $\eta_{k}(t)$ of the occupation 
numbers. A full update cycle $\eta(t) \to \eta(t+2)$ consists of two consecutive 
time steps. In the first step $t\to t+1$, a particle on site $2k-1$ in an odd bond
$(2k-1,2k)$ moves from $2k-1$ to $2k$, provided that site $2k$ is empty. 
Otherwise nothing happens in bond $(2k-1,2k)$. This jump rule is applied to all 
odd bonds simultaneously, corresponding to a deterministic sublattice
version of totally asymmetric random hopping like e.g. in the standard 
continuous-time TASEP \cite{Ligg99,Schu01}. In the second part $t+1\to t+2$ 
of an update cycle the same rule is applied to the even bonds $(2k,2k+1)$ 
except for the blockage bond $(L,1)$ on which a particle on site $L$ jumps 
randomly to site 1 with probability $p$ provided that site $1$ is 
empty.\footnote[1]{We mention that the sublattice property of the update 
dynamics has an equivalence with parallel update schemes without sublattice 
structure \cite{Raje98}.}

The invariant measure for a canonical ensemble of $N$ particles was derived in 
\cite{Schu93a} in terms of a set of selection rules and probability ratios
$p/q$. It was shown that at the critical density $\rho_c = p/2$ there is a 
phase transition in the thermodynamic limit $L\to\infty$ from a free flow 
phase for $\rho<\rho_c$ to a congested phase for $\rho>\rho_c$ with two
coexisting regimes of different densities, as described in the introductory 
discussion in the context of phase separation. Later a grandcanonical invariant 
measure -- where the conserved total particle number is a random variable -- 
was obtained in \cite{Hinr97} from a matrix product ansatz (MPA) \cite{Blyt07}, 
but not further investigated for this process. 

This matrix product approach is used in the present treatment to study the 
defect-induced nonequilibrium phase transition to the phase separated state 
rigorously and in considerably more detail than previously. In section 
\ref{Sec:MPA} we express the invariant measure in a matrix product form
similar to that of \cite{Hinr97} and point out the presence of a long-range 
{\it effective interaction} in the stationary distribution. Further properties are 
presented, discussed, and proved in sections \ref{Sec:phasetransition} 
(nonequilibrium phase transition), \ref{Sec:densityprofiles} (density profiles), 
and \ref{Sec:correlations} (correlation functions) where a long-range {\it 
correlation} resulting from the long-range effective interaction is explored.
In the appendices we list the properties of various 
functions used in the proofs (\ref{App:lim}) and we show how the matrix 
product representation of section \ref{Sec:MPA} follows from the MPA 
established in \cite{Hinr97} (\ref{App:MPA}).

A remark on the presentation: All mathematical results are exact. 
Their derivation is either elementary -- based solely on matrix 
multiplications and evaluations of geometrical series -- or uses well-established
properties of convergence of slowly varying discrete functions to continuous 
functions. No probabilistic or further advanced mathematical concepts are used. 
However, these derivations are lengthy, involve many case distinctions, 
and require precise statements concerning the range of validity of various 
mathematical functions appearing in the treatment. For clarity, we have 
therefore opted in most sections for an explicit separation between a statistical 
physics discussion of the results and their mathematical presentation in 
form of theorems and propositions which are followed by essentially 
rigorous computational proofs. 
 
\section{Stationary matrix product measure}
\label{Sec:MPA}

With the {\it i.i.d.} random variables $\zeta(t)$ with bimodal distribution
$f(\cdot) = (1-p) \delta_{\cdot,0} + p \delta_{\cdot,1}$ the dsTASEP 
described informally above is defined for $t\in\N_0$ by the update rules
\begin{eqnarray}
\left. \begin{array}{ll}
\displaystyle \eta_{2k-1}(t+1) & = \eta_{2k-1}(t) \eta_{2k}(t) \\
\displaystyle \eta_{2k}(t+1) & = 1 - \bar{\eta}_{2k-1}(t) \bar{\eta}_{2k}(t)
\end{array}
\quad  \right\} \quad t \mbox{ even}, \quad 1\leq k \leq \frac{L}{2} 
\end{eqnarray}
and
\begin{eqnarray}
\left. \begin{array}{ll}
\displaystyle \eta_{2k}(t+1) & = \eta_{2k}(t) \eta_{2k+1}(t) \\
\displaystyle \eta_{2k+1}(t+1) & = 1 - \bar{\eta}_{2k}(t) \bar{\eta}_{2k+1}(t)
\end{array}
\quad  \right\} \quad t \mbox{ odd}, \quad 1\leq k \leq \frac{L}{2}-1 
\nonumber \\
\left. \begin{array}{ll}
\displaystyle \eta_{L}(t+1) & = \eta_{L}(t) 
\left[1- \xi(t+1) \bar{\eta}_{1}(t)\right] \\
\displaystyle \eta_{1}(t+1) & = \eta_{1}(t) 
+ \xi(t+1) \bar{\eta}_{1}(t) \eta_{L}(t)
\end{array}
\quad  \right\} \quad t \mbox{ odd}.
\end{eqnarray}
In terms of the instantaneous currents
\begin{eqnarray}
j_{2k-1}(t) := \eta_{2k-1}(t) \bar{\eta}_{2k}(t), \qquad 
1 \leq k \leq \frac{L}{2} 
\label{instcurrodd} \\
j_{2k}(t) := [1 - \bar{\eta}_{2k-1}(t) \bar{\eta}_{2k}(t)] 
[1 - \eta_{2k+1}(t) \eta_{2k+2}(t)], \quad 1 \leq k \leq \frac{L}{2}-1 
\label{instcurreven} \\
j_{L}(t) := \zeta(t) [1 - \bar{\eta}_{L-1}(t) \bar{\eta}_{L}(t)] 
[1 - \eta_{1}(t) \eta_{2}(t)] ,
\label{instcurrL}
\end{eqnarray}
a full two-step update cycle is therefore expressed by the discrete continuity
equation
\begin{eqnarray}
\eta_{k}(t+2)  
& = & \eta_{k}(t) + j_{k-1}(t) - j_{k}(t), \qquad k\in\T_L, t\in \N_0
\end{eqnarray}
with the definition $j_{0}(t) := j_{L}(t)$. 

Under this jump dynamics the total particle number
\begin{equation}
N(\eta) = \sum_{k=1}^L \eta_k
\label{N}
\end{equation}
is conserved, but not the sublattice particle numbers
\begin{equation}
N^\pm(\eta) = \half \sum_{k=1}^L (1+(-1)^k) \eta_k .
\label{Nsl}
\end{equation}
The process is invariant under the particle-hole reflection symmetry 
$\eta_k \mapsto \bar{\eta}_{L+1-k}$ applied jointly to all $k$. 
We take $M=L/2$ even and focus on configurations $\eta$ with
$0 \leq N \leq L/2$ particles, corresponding 
to density $N/L \leq 1/2$. The properties of the model for $N/L > 1/2$ 
follow straightforwardly from the particle-hole symmetry. 

For $p=0$, particles cannot jump from site $L$ to site 1, corresponding to 
the trivial case of complete blockage where after a finite number of time 
steps all $N$ particles of a configuratipon $\eta$ pile up on the block of sites
$L-N+1, \dots , L$. Also for $p=1$ (no blockage) the dsTASEP becomes trivial 
after a finite number of steps as it reduces to deterministic translations of all
particles by one site per time step. Hence we restrict ourselves to the 
non-trivial range $0<p<1$ of the blockage parameter where translation 
invariance of the dynamics is broken.

To study the model in a grandcanonical ensemble we slightly modify the matrix 
product ansatz for the invariant measure developed in \cite{Hinr97}. To this end, we define 
the two-dimensional matrices
\begin{equation}
D := \frac{1}{2} \left(\begin{array}{cc}
1 & 1 \\ 1 & 1
\end{array}\right)
\label{MPAvectors}
\end{equation}
and 
\begin{equation}
A_0 := p \left(\begin{array}{cc}
0 & 1 \\ 0 & 1
\end{array}\right), \qquad 
A_1 := p \mathds{1}, \qquad 
A_2 := (1-p) \left(\begin{array}{cc}
1 & 1 \\ 0 & 0
\end{array}\right). 
\label{MPAmatrices}
\end{equation}
Furthermore, for $z\in\R$ we define
\begin{equation}
A := A_0 + z (A_1+A_2) =  \left(\begin{array}{cc}
z &  p + z (1-p) \\ 0 & p + p z
\end{array}\right) .
\end{equation}

With these matrices and the function
\begin{equation}
Y_K(p,z) := \Tr (D A^K)
\label{YKdef}
\end{equation}
the MPA of \cite{Hinr97} becomes
\begin{eqnarray}
P_{L,p,z}(\eta) 
& = & \frac{1}{Y_{\frac{L}{2}}(p,z)} \Tr \left\{ D
\left[\bar{\eta}_1\bar{\eta}_L A_0 + z \eta_1\bar{\eta}_L A_1
+ z \bar{\eta}_1 \eta_L A_2 \right] \right. \nonumber \\
& & \times \bar{\eta}_{2}
\left[\bar{\eta}_{L-1} A_0 + z \eta_{L-1} (A_1+A_2)\right]\nonumber \\
& & \times \dots \nonumber \\
& & \times
(\bar{\eta}_{2k-1}\bar{\eta}_{L+2-2k} A_0 
+ z \eta_{2k-1}\bar{\eta}_{L+2-2k} A_1
+ z \bar{\eta}_{2k-1} \eta_{L+2-2k} A_2 )\nonumber \\
& & \times \bar{\eta}_{2k}
\left[\bar{\eta}_{L+1-2k} A_0 + z \eta_{L+1-2k} (A_1+A_2)\right] 
\nonumber \\
& & \times \dots \nonumber \\
& & \times
(\bar{\eta}_{\frac{L}{2}-1}\bar{\eta}_{\frac{L}{2}+2} A_0 
+ z \eta_{\frac{L}{2}-1}\bar{\eta}_{\frac{L}{2}+2} A_1
+ z \bar{\eta}_{\frac{L}{2}-1} \eta_{\frac{L}{2}+2} A_2 )\nonumber \\
& & \left. \times \bar{\eta}_{\frac{L}{2}}
\left[\bar{\eta}_{\frac{L}{2}+1} A_0 
+ z  \eta_{\frac{L}{2}+1} (A_1+A_2)\right] \right\}.
\label{MPA}
\end{eqnarray}
We say that the measure $P_{L,p,z}(\eta)$ is a stationary 
matrix product measure 
(SMPM). The normalization factor $Y_{\frac{L}{2}}(p,z)$ plays the role of a 
grandcanonical nonequilibrium partition function in which
the total particle number $N(\eta)$ has a distribution determined by the 
parameter $z$ as can be seen by noting that 
$\sum_{\eta} z^{N(\eta)} P_{L,p,1}(\eta) = 
Y_{\frac{L}{2}}(p,z)/Y_{\frac{L}{2}}(p,1)$.
Thus it becomes evident that 
$z$ plays the role of a fugacity. Below we drop the
dependence of the SMPM on the blockage parameter $p$ and
the fugacity $z$.

One notices in the structure of the SMPM a fundamental difference between 
the region to the right of the blockage and the region to the left. To capture 
this phenomenon it is convenient to introduce lattice sectors.
\begin{definition}
A site $k\in\T_L$ is said to belong to sector 1, denoted by $\T_{L,1}$, if 
$k\in\{1,\dots,L/2\}$ and to sector 2, denoted by $\T_{L,2}$, if 
$k\in\{L/2+1,\dots,L\}$.
\end{definition}

Some other properties of the invariant measure that can be read off 
directly from of the structure of the SMPM \eref{MPA} and have analogs 
already found in \cite{Schu93a} in terms of a set of rules for the canonical 
ensemble with fixed particle number $N$. We generalize these rules here to 
the grandcanonical case.

\begin{proposition}
\label{Prop:smpm}
For any measurable function $f:\{0,1\}^L\to \R$ the SMPS has the projection 
properties
\begin{eqnarray}
\exval{\eta_{k} \eta_{L+1-k} f}_L = 0 & \qquad k \in \T_{L}
\label{smpm2} \\
\exval{\eta_{2k} f}_L = 0 & \qquad 2k \in \T_{L,1}
\label{smpm1a}
\end{eqnarray}
where $\exval{f}_L$ denotes the expectation of a function $f(\eta)$ 
w.r.t. \eref{MPA}.
\end{proposition}

\begin{remark}
The projection property \eref{smpm2} demonstrates that the invariant measure 
incorporates an long-range effective interaction between a site $k$ in the left 
segment $\T_{L,1}$ and the reflected site $L+1-k$ in the right segment 
$\T_{L,2}$, no matter how far (in lattice units) the two sites 
are apart. 
\end{remark}

The appearance of a stationary effective long-range interaction is somewhat 
counterintuitive since the microscopic dynamics is one-dimensional, completely 
local and has finite local state space. An immediate consequence are long-range
anticorrelations $\exval{\eta_k\eta_{L+1-k}}_L - 
\exval{\eta_k}\exval{\eta_{L+1-k}}_L = -
\exval{\eta_k}\exval{\eta_{L+1-k}}_L$. A long-range reflection property of 
correlations reminiscent of this anticorrelation was conjectured for the 
continuous-time TASEP with blockage \cite{Jano92}. We also find it intriguing 
that the SMPM is similar to a class of probability distributions for annihilating 
random walks \cite{Schu95}.

For explicit computations one needs to know the normalization $Y_{L/2}$.
In terms of the {\it critical fugacity}
\begin{equation}
\label{zc}
z_c := \frac{p}{1-p} 
\end{equation}
the $K^{\mathrm{th}}$ power of the matrix $A$ can be written as
\begin{eqnarray}
A^K & = & \left\{
\begin{array}{ll}
\displaystyle \left(\begin{array}{cc}
\displaystyle z^{K} &  \frac{z+z_c}{z-z_c} 
\left(z^{K} - p^{K} (1+z)^{K}\right)\\ 
\displaystyle 0 & p^{K} (1 + z)^{K}
\end{array}\right) & \qquad z \neq z_c \\[4mm]
p^{K} (1-p)^{-K} \left(\begin{array}{cc}
1 &  2 (1-p) K\\ 
0 & 1
\end{array}\right) & \qquad z = z_c .
\end{array}
\right.
\label{AM}
\end{eqnarray}
which is proved easily by induction. Therefore,
\begin{equation}
\label{YL}
Y_{K} =\cases{ \frac{z_c^{K+1}}{z_c-z} \left[\left(\frac{1+z}{1+z_c}\right)^K - 
\left(\frac{z}{z_c}\right)^{K+1}\right]
& $z \neq z_c$ \\
z_c^K [(1-p) K + 1] & $z = z_c$}.
\end{equation}
\normalsize
We point out that the limit $z\to z_c$ and the thermodynamic limit $L\to\infty$ 
may not commute in expectation values.

Furthermore, we recall the quadratic relations \cite{Hinr97}
\begin{eqnarray}
A_0^2 = p A_0, \qquad A_2^2 = (1-p) A_2
\label{Aqr1} \\
A_0 A_2 = 0
\label{Aqr3} \\
A_0 A = p (1+z) A_0, \qquad A A_2 = z A_2
\label{Aqr2} \\
 A_0 D = p D , \qquad D A_2 = (1-p) D .
\label{Aev}
\end{eqnarray}
From \eref{Aqr1} and \eref{Aqr3} together with the trivial relations 
$A_\alpha A_1 = A_1 A_\alpha = p A_\alpha$ one obtains the reduction
formula
\begin{equation}
(A_0+zA_1)(A_1+A_2) = p A.
\label{pA}
\end{equation}
Iterating the quadratic relations \eref{Aqr2} - \eref{Aev} yields
for $n\in\N_0$
\begin{eqnarray}
D A^n A_2 & = (1-p) z^n  D 
\label{Lem1b}
 \\
A_0 A^n D & = p^{n+1} (1+z)^n D 
\label{Lem1c} 
\end{eqnarray}
From \eref{Lem1b} one reads off the commutator property
\begin{eqnarray}
D A^n A_2 A_0 & = D A^n [A_2,A], \quad n\in\N_0 .
\label{Lem1d}
\end{eqnarray}
These matrix identities, in particular the reduction formulas \eref{pA},
\eref{Lem1b}, and the commutator property \eref{Lem1d}, will be 
used frequently in computations below. 
The quadratic relation \eref{Aqr3} leads to a further long range 
effective interaction inside sector $\T_{L,2}$ as it implies for any
measurable function $f$ the projection 
property 
\begin{eqnarray}
\exval{\eta_{2k} \bar{\eta}_{2k+2p-1} f}_L = 0, 
& \qquad 2k \in \T_{L,2}, \quad 1 \leq p \leq L/2-k 
\label{smpm1b}
\end{eqnarray}
noticed in \cite{Schu93a} for the canonical ensemble.

\section{Particle number fluctuations and stationary current}
\label{Sec:phasetransition}

The dynamics conserves the particle number, but the matrix product
measure is a mixture of canonical invariant measures with
particle number $N$ that, as shown below, has a non-trivial distribution
as a function of the blockage parameter $p$ and the fugacity $z$.
In particular, it turns out that there is a critical density below which
the variance of the particle number is proportional to the system size
$L$ -- corresponding to a non-zero thermodynamic compressibility --
while above the critical density there is a phase separated regime
where the variance reaches a constant for $L\to\infty$ so that the 
thermodynamic compressibility vanishes. This implies that
the two coexisting phases are $\it not$ subcritical bulk phases at two
different densities, as one might expect from equilibrium
phase separation e.g. in the two-dimensional Ising model.
Also the stationary current changes it behaviour at the critical point.

\subsection{Critical point and density fluctuations}

It was shown in \cite{Schu93a} for the canonical ensemble that a 
non-equilibrium phase transition occurs at a critical density $\rho_c = p/2$. 
Here we establish an analogous result $\rho_c := \rho(p,z_c) = p/2$
for the grandcanonical SMPM (\ref{MPA}) in terms of the critical fugacity $z_c$
(\ref{zc}) and discuss in detail the variance of the particle number.

\begin{theorem}
\label{Theo:density}
The particle density $\rho(p,z)$ 
\begin{eqnarray}
\rho(p,z) & := & \lim_{L\to\infty} \frac{1}{L} \exval{N}_L 
\label{Def:rho}
\end{eqnarray}
has a jump discontinuity at the critical point given by
\begin{eqnarray}
\rho(p,z) & = & \left\{ \begin{array}{lcl}
\displaystyle \frac{1}{2} \frac{z}{1+z}  & \quad & z < z_c \\[4mm]
\displaystyle \frac{1+p}{4} & & z = z_c \\[4mm]
\displaystyle \frac{1}{2} & & z \geq z_c .
\end{array}
\right. 
\label{rho} 
\end{eqnarray}
\end{theorem}

\begin{theorem}
\label{Theo:rhovariance}
The compressibility $C(p,z)$ 
\begin{eqnarray}
C(p,z) & := & \lim_{L\to\infty} \frac{1}{L} 
\left(\exval{N^2}_L - \exval{N}_L^2\right)
\label{Def:rhovariance}
\end{eqnarray}
diverges at the critical point and is given by
\begin{eqnarray}
C(p,z) & = & \left\{ \begin{array}{lcl}
\displaystyle  \frac{1}{2} \frac{z}{(1+z)^2} & \quad & z < z_c \\[4mm]
\displaystyle \infty & & z = z_c \\[4mm]
\displaystyle 0 & & z > z_c .
\end{array}
\right.
\label{rhovariance}
\end{eqnarray}
Moreover, for the critical regime $z\geq z_c$ one has
\begin{eqnarray}
& & \lim_{L\to\infty} \frac{1}{L^2} \left(\exval{N^2}_L - \exval{N}_L^2\right)
= \frac{(1-p)^2}{48}, \quad z = z_c,
\label{rhovariance2} \\
& & \lim_{L\to\infty} \left(\exval{N^2}_L - \exval{N}_L^2\right)
= \frac{z z_c}{(z-z_c)^2} \quad z > z_c.
\label{rhovariance3}
\end{eqnarray}
\end{theorem}

\begin{remark}
\label{Rem:1}
The supercritical particle variance (\ref{rhovariance3}) is also the amplitude of 
the finite-size correction to $C(p,z)$ for $z < z_c$ to leading order in $1/L$.
\end{remark}

\proof Both theorems are naturally proved together. For notational simplicity we 
suppress the dependence on $p$ and $z$ in all functions considered below.

Since the expectation of the total particle number 
can be written
\begin{eqnarray}
\exval{N}_L & = & \sum_{k=1}^{M} \exval{\eta_k+\eta_{L+1-k}}_L, \end{eqnarray}
the SMPM yields
\begin{eqnarray}
\exval{N}_L & =  z \frac{\mathrm{d}}{\mathrm{d} z} \ln{Y_{\frac{L}{2}}}
\label{NL1}
\end{eqnarray}
Moreover, from \eref{smpm2} in Proposition
\ref{Prop:smpm} one gets
$\exval{(\eta_k + \eta_{L+1-k})^2}_L = \exval{\eta_k+\eta_{L+1-k}}_L$ and
it follows that
\begin{equation}
\label{CL1}
C_L := \left(z \frac{\mathrm{d}}{\mathrm{d} z}\right)^2 \ln{Y_{\frac{L}{2}}}
= \frac{1}{Y_{\frac{L}{2}}} 
\left(z \frac{\mathrm{d}}{\mathrm{d} z}\right)^2 Y_{\frac{L}{2}}
- \left(\frac{1}{Y_{\frac{L}{2}}} z \frac{\mathrm{d}}{\mathrm{d} z}
 Y_{\frac{L}{2}}\right)^2 
\end{equation}
is the variance of the particle number in a finite system of length $L$.

For computing the derivatives w.r.t. $z$ and then taking the thermodynamic
limit it is convenient to introduce
\begin{equation}
\tilde{Y}_M := z_c^{-M} Y_{M}
= \left(\frac{z}{z_c}-1\right)^{-1}
\left[\left(\frac{z}{z_c}\right)^{M+1} - 
\left(\frac{1+z}{1+z_c}\right)^M\right].
\end{equation}
so that one can replace $p^{-M} Y_{M}$ in (\ref{NL1}) and (\ref{CL1})
by $\tilde{Y}_M$. 
One obtains
\begin{eqnarray}
z \frac{\mathrm{d}}{\mathrm{d} z} \ln{\tilde{Y}_M} & =  \frac{z}{z_c} 
\left(1-\frac{z}{z_c}\right)^{-1} \nonumber \\
&  + \frac{(M+1) \left(\frac{z}{z_c}\right)^{M+1} 
- M \frac{z}{1+z} \left(\frac{1+z}{1+z_c}\right)^M}
{\left(\frac{z}{z_c}\right)^{M+1} 
- \left(\frac{1+z}{1+z_c}\right)^M} .
\end{eqnarray}
Setting $M=L/2$ and taking the limit $L\to\infty$ immediately gives (\ref{rho}) 
for $z \neq z_c$.

Next we consider the density fluctuations. 
The second term in the second equality in \eref{CL1} is $\exval{N}_L^2$ and 
it remains to
compute the first term. Taking the derivative 
and dividing by $\tilde{Y}_M$ gives
\begin{eqnarray}
\frac{1}{\tilde{Y}_M} \left(z \frac{\mathrm{d}}{\mathrm{d} z}\right)^2 \tilde{Y}_M 
& = 2 \left(\frac{z}{z_c}\right)^2
\left(1-\frac{z}{z_c}\right)^{-2} + \left(\frac{z}{z_c}\right)
\left(1-\frac{z}{z_c}\right)^{-1} \nonumber \\
& + 2 \frac{z}{z_c} \left(1-\frac{z}{z_c}\right)^{-1} 
\frac{M \frac{z}{1+z} \left(\frac{1+z}{1+z_c}\right)^{M} 
- (M+1) \left(\frac{z}{z_c}\right)^{M+1}}{\left(\frac{1+z}{1+z_c}\right)^M 
- \left(\frac{z}{z_c}\right)^{M+1}} \nonumber \\
& + \frac{\left(\frac{M z}{(1+z)^2} + \frac{M^2 z^2}{(1+z)^2}\right) 
\left(\frac{1+z}{1+z_c}\right)^{M} 
- (M+1)^2 \left(\frac{z}{z_c}\right)^{M+1}}{\left(\frac{1+z}{1+z_c}\right)^M 
- \left(\frac{z}{z_c}\right)^{M+1}} 
\label{aux1b}
\end{eqnarray}
and therefore, after some rearrangement of terms,
\begin{eqnarray}
C_{L}
& = \half \frac{\frac{z}{(1+z)^2}}{1 - \frac{z}{z_c}
\left(\frac{z(1+z_c)}{z_c(1+z)}\right)^{L/2}}
+ \frac{1}{L} \frac{z z_c}{(z-z_c)^2} \nonumber \\
& - \frac{1}{L} \frac{\left(\frac{L}{2(1+z)}+1\right)^2 
 \frac{z}{z_c}
\left(\frac{z(1+z_c)}{z_c(1+z)}\right)^{L/2}}{\left(1 - \frac{z}{z_c}
\left(\frac{z(1+z_c)}{z_c(1+z)}\right)^{L/2}\right)^2} .
\label{CLfinite}
\end{eqnarray}
Taking the thermodynamic limit one arrives at (\ref{rhovariance}) for $z\neq 
z_c$. In the same way, by taking appropriate limits, one obtains 
(\ref{rhovariance3}) and the claim made in Remark \ref{Rem:1}.

To study the critical point we set $z = z_c(1+\epsilon)$ so that
\begin{eqnarray}
\ln{\tilde{Y}_M} 
& = \ln{\left[\left(1 + \epsilon\right)^{M+1} 
- \left(1 + \epsilon p\right)^M\right]} - \ln{\epsilon} \nonumber \\
& = \ln{\left[\sum_{n=0}^{M} {M+1 \choose n+1} \epsilon^{n}
- p \sum_{n=0}^{M-1} {M \choose n+1} (\epsilon p)^{n} \right]} 
\end{eqnarray}
and
\begin{equation}
z \frac{\rmd}{\rmd z} = (1+\epsilon) \frac{\rmd}{\rmd \epsilon} .
\end{equation}
Expanding in $\epsilon$ gives to the required second order
\begin{eqnarray}
\ln{\tilde{Y}_M} 
& = \ln{(M+1-pM)} + a_M \epsilon +\frac{\epsilon^2}{2} (2 b_M  - a_M^2) + 
O(\epsilon^3)
\end{eqnarray}
with
\begin{eqnarray}
a_M & = \frac{{M+1 \choose 2} - p^2 {M \choose 2}}{(1-p)M + 1} \\
b_M & = \frac{{M+1 \choose 3} - p^3 {M \choose 3}}{(1-p)M + 1} 
\end{eqnarray}
Therefore
\begin{eqnarray}
(1+\epsilon) \frac{\rmd}{\rmd \epsilon} \ln{\tilde{Y}_M} 
& = a_M + O(\epsilon) \\
\left((1+\epsilon) \frac{\rmd}{\rmd \epsilon} \right)^2\ln{\tilde{Y}_M} 
& = a_M + 2 b_M - a_M^2 + O(\epsilon)
\end{eqnarray}
which yields
\begin{equation}
\exval{N}_L = a_{\frac{L}{2}}, \qquad C_L = a_{\frac{L}{2}}
+ 2 b_{\frac{L}{2}} - a_{\frac{L}{2}}^2.
\end{equation}
Taking the thermodynamic limit for $\epsilon=0$ yields (\ref{rho}) and
(\ref{rhovariance}) for $z=z_c$ as well as (\ref{rhovariance2}). \qed

\subsection{Stationary current}

The stationary current is the space-independent expectation 
\begin{eqnarray}
j_L(p,z) & = & \exval{\eta_{2k-1} \bar{\eta}_{2k}}_L, \quad 1 \leq k \leq L/2 \\
& = & \exval{[1 - \bar{\eta}_{2k-1} \bar{\eta}_{2k}] 
[1 - \eta_{2k+1} \eta_{2k+2}]}_L,
\quad 1 \leq k < L/2 \\
& = & p \exval{(1 - \bar{\eta}_{L-1} \bar{\eta}_{L}) 
(1 - \eta_{1} \eta_{2})}_L
\end{eqnarray}
of the instantaneous currents (\ref{instcurrodd}) - (\ref{instcurrL}).

\begin{theorem}
\label{Theo:curr}
The macroscopic current $j(p,z) := \lim_{L\to\infty} j_L(p,z)$ is continuous
at the critical point and given by
\begin{equation}
\label{j}
j(p,z) = \cases{
\frac{z}{1+z} & $z < z_c$ \\
p & $z \geq z_c$ .
}
\end{equation}
\end{theorem}

\proof One obtains from the MPA (\ref{MPA}) for 
$\exval{\eta_{2k-1} \bar{\eta}_{2k}}_L$ in the range $1 \leq k \leq L/4$ 
\begin{eqnarray}
j_L(p,z)
& = & \frac{1}{Y_{\frac{L}{2}}} \sum_{\eta} 
\Tr \left\{ D
\left[\bar{\eta}_1\bar{\eta}_L A_0 + z \eta_1\bar{\eta}_L A_1
+ z \bar{\eta}_1 \eta_L A_2 \right] \right. \nonumber \\
& & \times \bar{\eta}_{2}
\left[\bar{\eta}_{L-1} A_0 + z \eta_{L-1} (A_1+A_2)\right]\nonumber \\
& & \times \dots \nonumber \\
& & \times 
z \eta_{2k-1}\bar{\eta}_{L+2-2k} A_1 \nonumber \\
& & \times \bar{\eta}_{2k}
\left[\bar{\eta}_{L+1-2k} A_0 + z \eta_{L+1-2k} (A_1+A_2)\right] \nonumber \\
& & \times \dots \nonumber \\
& & \times
(\bar{\eta}_{\frac{L}{2}-1}\bar{\eta}_{\frac{L}{2}+2} A_0 
+ z \eta_{\frac{L}{2}-1}\bar{\eta}_{\frac{L}{2}+2} A_1
+ z \bar{\eta}_{\frac{L}{2}-1} \eta_{\frac{L}{2}+2} A_2 )\nonumber \\
& & \left. \times \bar{\eta}_{\frac{L}{2}}
\left[\bar{\eta}_{\frac{L}{2}+1} A_0 + z \eta_{\frac{L}{2}+1} (A_1+A_2)\right] \right\}
\nonumber \\
& = & \frac{1}{Y_{\frac{L}{2}}} \sum_{\eta_{2k-1}} 
\sum_{\eta_{L+2-2k}} 
\Tr \left\{ D A^{2k-2}
z \eta_{2k-1}\bar{\eta}_{L+2-2k} A_1 A^{\frac{L}{2}-2k+1}\right\}
\nonumber \\
& = & zp \frac{Y_{\frac{L}{2}-1}}{Y_{\frac{L}{2}}}  .
\end{eqnarray}
For any nonnegative integer $K$ the normalization ratio is obtained 
from \eref{YL} as
\begin{equation}
\label{YMN}
\frac{Y_K}{Y_{\frac{L}{2}}} = \left\{ 
\begin{array}{ll}
\left[p\left(1+z\right)\right]^{K-\frac{L}{2}} 
\frac{1 - \frac{z}{z_c} \left(\frac{z(1+z_c)}{z_c(1+z)}\right)^{K}}
{1 - \frac{z}{z_c} \left(\frac{z(1+z_c)}{z_c(1+z)}\right)^{\frac{L}{2}}} 
& z \neq z_c \\[4mm]
z_c^{K-\frac{L}{2}} \frac{1 + (1-p) K}{1 + (1-p) \frac{L}{2}} & z = z_c .
\end{array}
\right.
\end{equation}
With the effective length
\begin{equation}
L_{\mathrm{eff}} := L + 2 (1-p)^{-1}
\label{Leffdef}
\end{equation}
one gets the exact result
\begin{eqnarray}
j_L(p,z) & = & \left\{ 
\begin{array}{ll}
\frac{z}{1+z}
\frac{1 - \frac{z}{z_c}
\left(\frac{z(1+z_c)}{z_c(1+z)}\right)^{L/2-1}}
{1 - \frac{z}{z_c} \left(\frac{z(1+z_c)}{z_c(1+z)}\right)^{L/2}} 
& z \neq z_c \\[4mm]
p \left(1 - \frac{2}{L_{\mathrm{eff}}}\right) & z = z_c .
\end{array}
\right.
\label{jL}
\end{eqnarray}
Taking the limit $L\to\infty$ yields (\ref{j}). \qed

\begin{remark}
The current $j(p,z)$ as function of the particle density is given by
\begin{equation}
\jrev(p,\rho) := j(p,z(\rho)) = \left\{ \begin{array}{ll}
\displaystyle 2 \rho & \mbox{ if } \rho < \rho_c \\
\displaystyle p & \mbox{ if } \rho \geq \rho_c 
\end{array}\right.
\end{equation}
which was observed already in \cite{Schu93a} for the canonical ensemble.
\end{remark}

\section{Sublattice density profiles}
\label{Sec:densityprofiles}

The density profile
\begin{equation}
\rho_L(k) := \exval{\eta_k}_L, \qquad k \in \T_L
\label{rhoLKdef}
\end{equation}
for the odd and even sublattices was computed in \cite{Schu93a} for the 
canonical ensemble. Here we consider the grandcanonical case and provide 
a full discussion of the limit $L\to\infty$. Guided by the canonical results. we 
introduce to this end also the shifted lattice defined by the set 
$\tilde{\T}_L = \{-L/2+1,\dots ,L/2\}$ of shifted lattice sites and 
occupation variables for nonpositive $k\in\{-L/2+1,\dots,0\}$ 
by $\eta_{k} = \eta_{k+L}$. We recall that $L$ is an integer multiple of
4 so that $L/2$ is even.

\subsection{Synopsis}

By equivalence of ensembles one expects in the free flow phase below the 
critical density similar results for the canonical and the grandcanonical 
measure \eref{MPA} when taking the thermodynamic limit. Indeed, as 
shown below, for both ensembles one has an essentially flat density profile 
except for a boundary layer to the left of the blockage inside sector 2 whose 
width is proportional to a constant $\xi$, i.e., does not grow as $L\to\infty$. 
In the domain wall picture of the density profile, the probability of finding 
the domain wall away from the blockage decays exponentially with parameter 
$\xi$. Therefore we call $\xi$ the {\it localization length}.

The behaviour in the grandcanonical ensemble at and above the critical point is 
different from the canonical case and clarified below. The linear density 
profile that we obtain at the critical point indicates that the domain wall 
position is unifomly distributed over the whole second lattice sector $\T_{L,2}$.
In contrast, in the canonical ensemble the domain wall is confined to a region
of size $\sqrt{L}$ to the left of the blockage inside sector 2. 

In the phase separated state above the critical point, the grandcanonical 
density profile has essentially two regions of homogeneous density, viz., of low 
density in sector 1 and of high density in sector 2, separated by a boundary 
layer to the right of the lattice center in sector 2. On the other hand, in the 
canonical ensemble the domain wall fluctuates inside the lattice segment 2 
around a mean position that is determined by the fixed excess density 
$\rho-\rho_c$. The fluctuations remain confined to a region of size $\sqrt{L}$ 
around the mean position.

\subsection{Exact results}

To state and further discuss these results in precise form we recall the 
definitions of the floor function
\begin{equation}
\floor{x} = \max{\{n\in\Z \, | \, n \leq x\}}, \qquad x\in \R
\label{floor}
\end{equation} 
and the Heaviside indicator function and its complement
\begin{equation}
\Theta_x = \cases{1 & for $x>0$ \\ 0 & for $x\leq 0$ }, \qquad 
\bar{\Theta}_x := 1-\Theta_x, \qquad x\in \R 
\label{Heaviside}
\end{equation}
where for $n\in\Z$ one has $\bar{\Theta}_{n} = \delta_{n,0} + 
\Theta_{-n}$.
We also introduce for $n \in \Z$ the sublattice indicator function
\begin{eqnarray}
Q^\pm_n & := \half \left(1 \pm (-1)^n\right).
\label{indisubdef}
\end{eqnarray}
and the open intervals
\begin{equation}
\I_1 := (0,\half), \qquad \I_2 = (\half,1)
\end{equation}
that are continuum analogs of the lattice sector sets $\T_{L,\alpha}$.

To avoid heavy notation, the dependence of most 
functions on the parameters $p$ and $z$ will be suppressed.

\subsubsection{Asymptotic sublattice density profiles on lattice scale}

To probe the bulk density we fix a reference position deep inside each
lattice sector and study the density profile around this bulk position in the 
thermodynamic limit.

\begin{theorem}[Offcritical bulk density]
\label{Theo:bulkdensity} Let $z\neq z_c$ be offcritical. The bulk density profile 
in sector $\alpha$ defined by
\begin{eqnarray}
\rho^{\mathrm{bulk}}_\alpha(k) := \lim_{L\to\infty} 
\exval{\eta_{2\floor{L u/2} + k}}_L , \qquad \mbox{for }u \in \I_\alpha, k\in\Z
\label{rhobulkdef}
\end{eqnarray}
depends only on the sublattice and is given by
\begin{eqnarray}
\rho^{\mathrm{bulk}}_1(k) = j Q^-_k, \qquad 
\rho^{\mathrm{bulk}}_2(k) = \rho^{\mathrm{bulk}}_1(k) + \cases{0 & $z<z_c$ \\ \frac{j}{z_c} & $z>z_c$}
\label{rhobulk} 
\end{eqnarray}
with the stationary current $j$ \eref{j}.
\end{theorem}

To see the boundary layers announced in the synopsis, the density profile 
in the thermodynamic limit $L\to\infty$ needs to be studied on lattice scale. This analysis has to be done separately around the blockage bond $(L,1)$ on the
one hand and around the central bond $(L/2,L/2+1)$ on the other hand.

\begin{theorem}[Boundary layer profiles]
\label{Theo:dpthermo}
Let $z\neq z_c$. In terms of the localization length 
\begin{equation}
\label{loclengthdef}
\xi := |\ln{\frac{z_c(1+z)}{z(1+z_c)}}|^{-1}
\end{equation}
and the boundary layer functions
\begin{eqnarray}
\sigma(k) & := \frac{j}{z_c} \rme^{-\frac{|k|}{\xi}} 
\label{dbldef} \\
\tilde{\sigma}(k) & := \frac{j}{z_c} \frac{1+z_c}{1+z} 
\rme^{-\frac{|k|}{\xi}} 
\label{cbldef}
\end{eqnarray}
with the critical fugacity $z_c$ \eref{zc} and the current $j$ \eref{j},
the density profile as seen from the blockage bond defined in the thermodynamic 
limit by
\begin{equation}
\label{denspro}
\rho_{\infty}(k) := \cases{\lim_{L \to \infty} \rho_L(k) & $k>0$ \\ 
\lim_{L \to \infty} \rho_L(L-|k|) & $k\leq 0$}
\end{equation}
and the central density profile
\begin{equation}
\label{densprocent}
\tilde{\rho}_{\infty}(k) := \lim_{L \to \infty} 
\rho_L(L/2+k)
\end{equation}
as seen in the thermodynamic limit from the central bond are given by
\begin{eqnarray}
\rho_{\infty}(k) & =
\cases{\rho^{\mathrm{bulk}}_1(k) + \sigma(k) \bar{\Theta}_{k} 
&  $z < z_c$ \vspace*{2mm}\\
\rho^{\mathrm{bulk}}_1(k) \Theta_{k} 
+ \rho^{\mathrm{bulk}}_2(k) \bar{\Theta}_{k} & $z > z_c$
} 
\label{rhok} \\[2mm]
\tilde{\rho}_{\infty}(k) & = \cases{\rho^{\mathrm{bulk}}_1(k) 
&  $z < z_c$ \vspace*{2mm} \\
\rho^{\mathrm{bulk}}_1(k) 
\bar{\Theta}_{k} +  \rho^{\mathrm{bulk}}_2(k) \Theta_{k}   - \tilde{\sigma}(k)
& $z > z_c$ \\
} 
\label{trhok}
\end{eqnarray}
with the 
bulk density functions $\rho^{\mathrm{bulk}}_\alpha(k)$ \eref{rhobulk} for 
all $k\in\Z$.
\end{theorem}


We point out that seen from the blockage bond, nonpositive values $k$ 
correspond to sector 2 of the lattice whereas seen from the central bond,
nonpositive values $k$ correspond to sector 1 of the lattice. 
Theorems \ref{Theo:bulkdensity} and \ref{Theo:dpthermo} thus
essentially assert that below the critical point the density profile is 
homogeneous with sublattice-dependent (but otherwise constant) amplitude 
$\rho^{\mathrm{bulk}}_1(k) = j Q^-_k$, except for a boundary layer 
\eref{dbldef}
to the left of the blockage inside sector 2, while above 
the critical point the density ``jumps'' on each sublattice at the central bond 
and at the blockage bond between
$\rho^{\mathrm{bulk}}_1(k)$ and $\rho^{\mathrm{bulk}}_2(k) =
1-p + \rho^{\mathrm{bulk}}_1(k)$, except for a central boundary layer
\eref{cbldef} inside sector 2 that interpolates to the right of the central bond 
between $\rho^{\mathrm{bulk}}_1(k)$ and $\rho^{\mathrm{bulk}}_2(k)$. 
%

%

\subsubsection{Critical density profile}

As one approaches the critical point, the localization length diverges and the 
notion of boundary layer looses its meaning. To explore the density profile at 
the critical point we employ a hydrodynamic scaling $L\to\infty$ with lattice 
sites seen from the center and taken as $k = \floor{uL}$ with constant 
$u \in (-1/2,1/2]$. Here $u$ has the meaning of a macroscopic position $u$ 
on a circle of unit length $\ell=1$ with the blockage at $u=1/2$.
Due to the finite sublattice alternation of the local density coming from the
term $j Q^-_k$ in the microscopic density profiles, this limit has to 
be taken separately for each sublattice. This is achieved by the
sublattice decomposition 
\begin{equation}
\rho_L(k) = \rho^+_L(k/2) Q^+_k + \rho^-_L(\floor{k/2}+1) Q^-_k
\end{equation}
 of the density profile with the sublattice density profiles
\begin{eqnarray}
\rho^+_L(k) := \rho_L(2k), \qquad
\rho^-_L(k) := \rho_L(2k-1), \qquad 1 \leq k \leq L/2 .
\end{eqnarray}
Analogously, and for reference, we define the supercritical sublattice 
bulk densities
\begin{eqnarray}
& \rho^{\mathrm{bulk},-}_{\mathrm{ps},1} := p, \qquad & \rho^{\mathrm{bulk},+}_{\mathrm{ps},1} := 0 \\
& \rho^{\mathrm{bulk},-}_{\mathrm{ps},2} := 1, & \rho^{\mathrm{bulk},+}_{\mathrm{ps},2} := 1-p 
\end{eqnarray}
obtained from \eref{rhobulk} for $z>z_c$.

\begin{theorem}[Macroscopic density profile]
\label{Theo:dphydro}
For $u \in (-1/2,1/2]$ the macroscopic sublattice density profiles 
under hydrodynamic scaling 
\begin{eqnarray}
\tilde{\rho}^{\pm}(u) := \lim_{L \to \infty}  \rho^\pm_L(L/4+\floor{uL/2}), 
\qquad 
\label{densprocrit} 
\end{eqnarray}
are piece-wise linear and given by
\begin{eqnarray}
\tilde{\rho}^{+}(u) & = \frac{j}{z_c} \Theta(u) \times \cases{
0  &  $z < z_c$ \\
2 u &  $z = z_c$ \\
1 & $z > z_c$} 
\\[4mm]
\tilde{\rho}^{-}(u) & = \tilde{\rho}^{+}_c(u) + j
\end{eqnarray}
with the macroscopic current $j$ of Theorem \ref{Theo:curr}.
\end{theorem}

One sees that at the critical point the boundary layer becomes ``infinitely''
wide in the sense that inside sector 2 it interpolates smoothly on each 
sublattice between the supercritical bulk densities 
$\rho^{\mathrm{bulk},\pm}_{\mathrm{ps},1}$ (everywhere inside sector 1) and 
$\rho^{\mathrm{bulk},\pm}_{\mathrm{ps},2}$ (attained only as one reaches the blockage)
that characterize the phase-separated state.

\subsection{Proofs}

All results follow from exact computation of the density profile
for the finite lattice using the SMPM \eref{MPA} and then taking the
limit $L\to\infty$ as defined in each theorem.

\proof We define for $n\in\{0,\dots,L/2\}$ the functions
\begin{equation}
H_L(n) := p(1-p) z^{n} \frac{Y_{\frac{L}{2}-n}}{Y_{\frac{L}{2}}}, \qquad
\tilde{H}_L(n) := H_L(L/2+1-n)
\label{HLdef}
\end{equation}
related to the ratio of partition functions \eref{YMN}. 
In particular, we note that \eref{Lem1b} yields
\begin{eqnarray}
\Tr (D A^{n-1} A_2 A^{\frac{L}{2}-n}) & 
= (1-p) z^{n-1} Y_{\frac{L}{2}-n} 
\label{Lem1b1} 
\end{eqnarray}
and therefore
\begin{eqnarray}
\frac{z}{Y_{\frac{L}{2}}} \Tr (D A^{n-1} A_2 A^{\frac{L}{2}-n})
& = \frac{1}{p}  H_L(n).
\label{Lem1b2} 
\end{eqnarray}

From Proposition \ref{Prop:smpm} one has for sector 1 
\begin{eqnarray}
\exval{\eta_{2k-1}}_L
& = p z \frac{Y_{\frac{L}{2}-1}}{Y_{\frac{L}{2}}} = j_L, \qquad
1\leq k \leq L/4
\label{rhoodd1}
\end{eqnarray}
while for sector 2 the SMPM yields
\begin{eqnarray}
\exval{\eta_{L+1-2k}}_L 
& = \frac{z}{Y_{\frac{L}{2}}} 
\Tr (D A^{2k-1} (A_1+A_2) A^{\frac{L}{2}-2k}) 
\nonumber \\
& = j_L + \frac{1}{p} H_L(2k), \qquad
1\leq k \leq L/4.
\label{etaL+1-2k}
\end{eqnarray}
In the last equality \eref{Lem1b2} was used. 

For even sites one has from Proposition \ref{Prop:smpm} for sector 1
\begin{eqnarray}
\exval{\eta_{2k}}_L & = & 0 , \qquad
1\leq k \leq L/4
\label{rhoeven1}
\end{eqnarray}
and for sector 2 one gets from the SMPM
\begin{eqnarray}
\exval{\eta_{L+2-2k}}_L
& = \frac{z}{Y_{\frac{L}{2}}} \Tr (D A^{2k-2} A_2 A^{\frac{L}{2}+1-2k})
\nonumber \\
& = \frac{1}{p} H_L(2k-1) , \qquad
1\leq k \leq L/4.
\label{etaL+2-2k}
\end{eqnarray}

The results \eref{rhoodd1} - \eref{etaL+2-2k}  can be written compactly as
\begin{eqnarray}
\rho_L(n) & = j_L Q^-_n + \frac{1}{p} H_L(L+1-n) \Theta^{(2)}_{L,n},
\qquad n \in \T_L
\label{rhoLn}
\end{eqnarray}
which yields
\begin{eqnarray}
\rho_L(L-n) & = j_L Q^-_n 
+ \frac{1}{p} H_L(n+1) \Theta^{(1)}_{L,n+1}, \qquad 0 \leq n \leq L-1 
\label{rhoLn2} \\
\rho_L(L/2+n) & = j_L Q^-_n 
+ \frac{1}{p} \tilde{H}_L(n) \Theta^{(1)}_{L,n},
\qquad -L/2+1 \leq n \leq L/2 
\label{trhoLn} \\
\rho^+_L(L/4+n) & = \frac{1}{p} \tilde{H}_L(2n-1) \Theta^{(1)}_{L,2n}
\qquad -L/4 < n \leq L/4 \\
\rho^-_L(L/4+n) & = j_L + \frac{1}{p} \tilde{H}_L(2n) \Theta^{(1)}_{L,2n-1}
\qquad - L/4 < n \leq L/4.
\end{eqnarray}
To take the thermodynamic limit $L\to\infty$ as indicated in each theorem 
one uses Theorem \ref{Theo:curr} for the current, the property
of the floor function $\floor{uL} = uL + R(u)$ with $0 \leq R(u) < 1$ 
uniformly bounded in $L$, and the asymptotic properties 
of the function $H_L(n)$ detailed in \ref{App:H}. \qed

\section{Correlations}
\label{Sec:correlations}

The local density $\exval{\eta_k}_L$ alone provides little information on 
the microscopic structure of the particle system. Studying correlations 
between the occupation numbers at two different sites
probes the role of the blockage in the formation of microscopic shocks
and yields insight into microsopic origin of the the particle number 
fluctuations in the grandcanonical ensemble.

Due to the absence of translation invariance, the density
correlation function
\begin{eqnarray}
S_L(k,l) & := \exval{\eta_{k} \eta_{l}}_L - \exval{\eta_{k}}_L 
\exval{\eta_{l}}_L  
\label{Skl} 
\end{eqnarray}
depends on both space coordinates $k,l$. By construction, 
\begin{equation}
S_L(l,k) = S_L(k,l)
\label{Slk}
\end{equation}
for all $k,l \in \T_L$. 

We define the dynamical structure function as the space average 
\begin{eqnarray}
S_L(r) & := \frac{1}{L} \sum_{k=1}^{L} S_L(k,(k+r)\ \mathrm{mod}\ L)
\label{SLrdef} 
\end{eqnarray}
of the two-point density correlation function \eref{Skl}. 
From the symmetry 
\eref{Slk} one deduces that $S_L(r) = S_L(-r)$.\footnote[2]{For translation
invariant lattice systems this definition reduces to the usual one
$S_L(r) = S_L(k,(k+r)\ \mathrm{mod}\ L)$ which is independent of $k$.}
To avoid heavy notation we omit the dependence on $(p,z)$ of the
functions appearing in this section.

\subsection{Density correlation function}

For the canonical ensemble some properties of the density correlation function 
\eref{Skl} were computed in \cite{Schu93a} with emphasis on the behaviour 
near the blockage. It was found that below the critical point the amplitude of 
correlations decays exponentially with parameter $\xi$ with 
increasing distance from the blockage while at the critical point there are
long-range correlations that extend over a region proportional to
$\sqrt{L}$. Here we provide a full discussion in the grandcanonical 
ensemble defined by the SMPM \eref{MPA}.

\subsubsection{Synopsis}

The main results concern the critical point 
and the phase separated regime: (i) For $z>z_c$ we identify short-range 
correlations near the center of the lattice that arise from the presence of 
the central boundary layer \eref{cbldef}. (ii) For $z\geq z_c$ we find a 
{\it long-range anticorrelation} between site $k$ and its reflected site 
$L+1-k$. (iii) At the critical point the system is shown to exhibit further 
long-range correlations with amplitude of order 1, extending over the 
whole sector 2. These correlations are indicative of a fluctuating 
microscopic shock as typical stationary configuration of the dsTASEP
in the grandcanonical ensemble, with the domain wall position uniformly
distributed over sector 2. These correlations are in contrast to those found 
for the canonical ensemble which extend only over a region of order 
$\sqrt{L}$ inside sector 2 and they also differ from those observed 
in the deterministic sublattice TASEP with open boundaries 
\cite{Schu93b,Hinr96,Jafa09} where the long-range correlation at criticality 
extends over the whole lattice and where the reflective contribution to the 
correlations is absent.

\subsubsection{Preparatory remarks and definitions}

On the diagonal $k=l$ in the $(k,l)$-plane the correlation function trivially 
has a non-vanishing term 
\begin{equation}
S^{\mathrm{hc}}_L(k,l) := A^{\mathrm{hc}}_L(k) \delta_{k,l}, \qquad
A^{\mathrm{hc}}_L(k) =\exval{\eta_{k}}_L - \exval{\eta_{k}}_L^2
\end{equation}
due to hard core exclusion. On the other hand, the reflective projection property 
\eref{smpm2} induces an non-trivial anticorrelation 
\begin{equation}
S^{\mathrm{refl}}_L(k,l) := A^{\mathrm{refl}}_L(k) \delta_{k+l,L+1}, \qquad
A^{\mathrm{refl}}_L(k) = - \exval{\eta_{k}}_L \exval{\eta_{L+1-k}}_L
\end{equation}
between site $k$ and the site $L+1-k$ reflected at the blockage bond $(L,1)$,
i.e., along the perpendicular diagonal $k=L+1-l$. Therefore we decompose the 
correlation function into the three parts
\begin{equation}
S_L(k,l) = A^{\mathrm{hc}}_L(k)\delta_{k-l,0} 
+ A^{\mathrm{refl}}_L(k) \delta_{k+l,L+1} + S^{\mathrm{bl}}_L(k,l)
\end{equation}
with the off-diagonal contribution
\begin{eqnarray}
S^{\mathrm{bl}}_L(k,l) & := S_L(k,l) 
\left(1 - \delta_{k-l,0} - \delta_{k+l,L+1}\right)
\end{eqnarray}
that, as it will turn out, has its origin in the boundary layers.
Similarly, we decompose the density correlation function in the 
thermodynamic limit defined for fixed values of $k,l \in \Z$ by
\begin{eqnarray}
S_{\infty}(k,l) & :=
\lim_{L\to\infty} \left[S_L(k,l) \Theta_{k} \Theta_{l} 
+ S_L(k,L-|l|) \Theta_{k} \bar{\Theta}_{l} \right. \nonumber \\
& \left. + S_L(L-|k|,l) \bar{\Theta}_{k} \Theta_{l} 
+ S_L(L-|k|,L-|l|) \bar{\Theta}_{k} \bar{\Theta}_{l} \right]
\label{Sthermo} \\[4mm]
\tilde{S}_{\infty}(k,l) & := \lim_{L\to\infty} S_L(L/2+k,L/2+l)
\label{tSthermo}
\end{eqnarray}
to study correlations around the blockage and around the lattice center 
respectively. With 
\begin{eqnarray}
A^{\mathrm{hc}}_{\infty}(k) & := \lim_{L\to\infty} 
A^{\mathrm{hc}}_{L}(k) \Theta_{k} + A^{\mathrm{hc}}_{L}(L-|k|) 
\bar{\Theta}_{k} \\
A^{\mathrm{refl}}_{\infty}(k) & := \lim_{L\to\infty} 
A^{\mathrm{refl}}_{L}(k) \Theta_{k} + A^{\mathrm{refl}}_{L}(L-|k|) 
\bar{\Theta}_{k} \\
\tilde{A}^{\mathrm{hc}}_{\infty}(k) & := \lim_{L\to\infty} A^{\mathrm{hc}}_{L}(L/2+k) \\
\tilde{A}^{\mathrm{refl}}_{\infty}(k) & := \lim_{L\to\infty}
A^{\mathrm{refl}}_{L}(L/2+k) \\
S^{\mathrm{bl}}_{\infty}(k,l) & :=
\lim_{L\to\infty} \left[S^{\mathrm{bl}}_L(k,l) \Theta_{k} \Theta_{l} 
+ S^{\mathrm{bl}}_L(k,L-|l|) \Theta_{k} \bar{\Theta}_{l} \right. \nonumber \\
& \left. + S^{\mathrm{bl}}_L(L-|k|,l) \bar{\Theta}_{k} \Theta_{l} 
+ S^{\mathrm{bl}}_L(L-|k|,L-|l|) \bar{\Theta}_{k} \bar{\Theta}_{l} \right]
\label{Rthermo} \\[4mm]
\tilde{S^{\mathrm{bl}}}_{\infty}(k,l) & := \lim_{L\to\infty} 
S^{\mathrm{bl}}_L(L/2+k,L/2+l)
\label{tRthermo}
\end{eqnarray}
the corresponding decompositions read
\begin{eqnarray}
S_{\infty}(k,l) & = A^{\mathrm{hc}}_{\infty}(k)\delta_{k-l,0} 
+ A^{\mathrm{refl}}_{\infty}(k) \delta_{k+l,1} + S^{\mathrm{bl}}_{\infty}(k,l) \\
\tilde{S}_{\infty}(k,l) & = \tilde{A}^{\mathrm{hc}}_{\infty}(k)\delta_{k-l,0} 
+ \tilde{A}^{\mathrm{refl}}_{\infty}(k) \delta_{k+l,1} 
+ \tilde{S}^{\mathrm{bl}}_{\infty}(k,l) 
\end{eqnarray}
for $k,l\in\Z$. We point out the symmetries
\begin{eqnarray}
A^{\mathrm{refl}}_{L}(k) 
= A^{\mathrm{refl}}_{L}(L+1-k) , \qquad
A^{\mathrm{refl}}_{\infty}(k) 
= A^{\mathrm{refl}}_{\infty}(1-k) \\
\tilde{A}^{\mathrm{refl}}_{L}(k) 
= \tilde{A}^{\mathrm{refl}}_{L}(1-k), \qquad
\tilde{A}^{\mathrm{refl}}_{\infty}(k) 
= \tilde{A}^{\mathrm{refl}}_{\infty}(1-k) 
\end{eqnarray}
that follow from the definitions of these quantities.

We also introduce the constant
\begin{eqnarray}
\kappa_L & := j_L(1-j_{L-2})
\label{kappaLdef} \\
& = \cases{
\frac{z}{(1+z)^2}\frac{1 - \left(\frac{1+z}{1+z_c}\right)^2 \rme^{-L/(2\xi_s)}}
{1 - \frac{z}{z_c} \rme^{-L/(2\xi_s)}} 
& $z \neq z_c$ \\
p(1 - p) - \frac{2p(1 - 2p)}{L_{\mathrm{eff}}}
& $z = z_c$} 
\end{eqnarray}
which has the limiting behaviour
\begin{eqnarray}
\kappa & := \lim_{L\to\infty} \kappa_L 
= \cases{\frac{z}{(1+z)^2}
 & $z < z_c$ \\
p (1-p) & $z \geq z_c$ } 
\label{kappa} 
\end{eqnarray}
Away from the critical point, finite-size corrections to the asymptotic result 
are exponentially small in $L$. For $z<z_c$ one has $\kappa = 2C$ with the 
subcritical compressibity $C$ \eref{rhovariance} established in Theorem 
\ref{Theo:rhovariance}.

\subsubsection{Main results}

As a reference, we begin with the offcritical bulk correlations. To this end, 
we fix inside the bulk of the sectors an arbitrary reference pair of lattice 
points $(m,n) = (2\floor{L u/2},2\floor{L v/2})$
and study the correlations in the thermodynamic limit at an 
arbitrary but finite distance around these points.

\begin{theorem}
\label{Theo:bulkcorrelation} 
For $z\neq z_c$ and fixed $k,l\in\Z$ the bulk correlations 
\begin{eqnarray}
S^{\mathrm{bulk}}_{\alpha\beta}(k,l) := \lim_{L\to\infty} 
S_L(2\floor{L u/2} + k, 2\floor{L v/2} + l)
\qquad u \in \I_{\alpha}, v \in \I_{\beta}
\label{Sabbulkdef}
\end{eqnarray}
are given by
\begin{eqnarray}
S^{\mathrm{bulk}}_{\alpha\beta}(k,l) = \cases{
A^{\mathrm{hc}}_{\alpha}(k) \delta_{k,l} & $\alpha=\beta, \quad u=v$ \\
A^{\mathrm{refl}}_{\alpha}(k) \delta_{k+l,1} 
& $\alpha \neq \beta, \quad u = 1 - v$ \\
0 & else 
}
\end{eqnarray}
with the sector-dependent bulk amplitudes
\begin{eqnarray}
A^{\mathrm{hc}}_{1}(k) = \kappa Q^-_k  \qquad z\neq z_c, \qquad
& A^{\mathrm{hc}}_{2}(k) = \kappa \cases{ 
Q^-_k & $z < z_c$ \\
Q^+_k & $z > z_c$ }
\label{S22bulk} \\
A^{\mathrm{refl}}_{1}(k) =  - \kappa \cases{
0 & $z < z_c$ \\
Q^-_k  & $z > z_c$
}, \quad
& A^{\mathrm{refl}}_{2}(k) = - \kappa \cases{ 
0 & $z < z_c$ \\
Q^+_k  & $z > z_c$
}
\label{S21bulk}
\end{eqnarray}
proportional to $\kappa$ as given in \eref{kappa}.
\end{theorem}

\begin{remark} In the free-flow phase below the critical point the
bulk correlations reduce to the hard-core onsite correlations with amplitude 
$\kappa=2C$ proportional to the compressibility \eref{rhovariance}. In the 
phase separated state above the critical point, the theorem asserts that in 
addition to the hard-core contribution there are bulk anticorrelations with 
negative amplitude proportional to $\kappa = p(1-p)$. 
These correlations are long ranged since the 
correlated occupation numbers $\eta_m$ and $\eta_n$ at the lattice points 
$m=2\floor{L u/2} + k$ and $n=2\floor{L (1- u)/2} + 1 - k$ 
have a nonzero {\it macroscopic} distance $r=|1-2u|$ in the 
thermodynamic limit.
\end{remark}

Next we investigate the offcritical correlations arising from the existence
of the boundary layers. We recall that seen from the blockage (center), sector 2 
(sector 1) of the finite lattice corresponds to negative lattice points in the 
thermodynamic limit.

\begin{theorem}[Offcritical correlations near the blockage]
\label{Theo:correlationsd} 
Seen from the blockage, the off-critical density correlation function has the 
hard-core part 
\begin{eqnarray}
A^{\mathrm{hc}}_{\infty}(k)
& = \cases{
\kappa Q_{k}^{-} +  \frac{\kappa}{z_c} \rme^{-(|k|/\xi} 
\left(1 + (-1)^k z - \frac{z}{z_c} \rme^{-|k|/\xi} \right) \bar{\Theta}_{k} 
& $z < z_c$ 
\vspace*{4mm} \\
\kappa \left[Q^-_{k} \Theta_{k} + Q^+_{k} \bar{\Theta}_{k}\right] 
& $z > z_c$,
} 
\label{Sklhcd}
\end{eqnarray}
the reflective anticorrelations
\small
\begin{eqnarray}
A^{\mathrm{refl}}_{\infty}(k)
& = \cases{- \frac{z_c}{(1+z_c)^2}
\rme^{-(|k-\half|+\half)/\xi} 
\left(Q^-_{k} \Theta_{k} +  Q^+_{k} \bar{\Theta}_{k} \right)
 & $z<z_c$ 
\vspace*{4mm} \\
- \kappa \left[Q^-_{k} \Theta_{k} + Q^+_{k} \bar{\Theta}_{k}\right]
& $z>z_c$,
}
\label{Sklrefld}
\end{eqnarray}
\normalsize
and the boundary layer correlations
%
\small
\begin{eqnarray}
S^{\mathrm{bl}}_{\infty}(k,l)
& = \cases{
\frac{z_c-z}{1+z_c} \frac{\kappa}{z_c} \bar{\Theta}_{k+l} 
\left[\rme^{-|k|/\xi} 
Q^-_{l} \bar{\Theta}_{k}\Theta_{l}
+ \rme^{-|l|/\xi} Q^-_{k} \Theta_{k} \bar{\Theta}_{l} \right]
& \\
+ \frac{\kappa}{z_c} \rme^{-|k|/\xi}   
\left[ \left(1 - \frac{z}{z_c} \rme^{-|l|/\xi} \right) 
Q^-_{l} \right.
& \\
\left. + \left(\frac{1+z}{1+z_c} - \frac{z}{z_c} \rme^{-|l|/\xi} \right)
Q^+_{l} \right] \Theta_{l-k} \bar{\Theta}_{k} \bar{\Theta}_{l} 
& \\
+ \frac{\kappa}{z_c} \rme^{-|l|/\xi}   
\left[ \left(1 - \frac{z}{z_c}\rme^{-|k|/\xi} \right) 
Q^-_{k} \right.
& \\
\left. + \left(\frac{1+z}{1+z_c} - \frac{z}{z_c}\rme^{-|k|/\xi} \right)
Q^+_{k} \right] \Theta_{k-l} \bar{\Theta}_{k} \bar{\Theta}_{l} 
& $z<z_c$ 
\vspace*{4mm} \\
0 & $z>z_c$ .
}
\label{Skl0d}
\end{eqnarray}
\normalsize
\end{theorem}

\begin{remark}
Above the critical point, the correlations are equal to the bulk values of 
Theorem \ref{Theo:bulkcorrelation} for all $k,l\in\Z$. Below the critical 
point, they come arbitrarily close to the bulk values as $\max{\{k,l\}}$ 
becomes large compared to the localization length $\xi$.
\end{remark}

\begin{theorem}[Offcritical correlations near the center]
\label{Theo:correlationsc} Seen from the center, the off-critical density 
correlation function has the hard core part
\small
\begin{eqnarray}
\tilde{A}^{\mathrm{hc}}_{\infty}(k)
= \cases{
\kappa Q_{k}^{-} & $z < z_c$
\vspace*{4mm} \\
\kappa \left[ Q^+_{k} \Theta_{k} + Q^-_{k} \bar{\Theta}_{k} \right] & \\
\frac{\rme^{-|k|/\xi}}{(1+z_c)(1+z)} 
\left(1 - z_c(-1)^k - \frac{1+z_c}{1+z} \rme^{-|k|/\xi} \right)
 \Theta_{k}
& $z > z_c$,
}
\label{TheoShcc}
\end{eqnarray}
\normalsize
the reflective contribution
\small
\begin{eqnarray}
\tilde{A}^{\mathrm{refl}}_{\infty}(k) = 
\cases{0 & $z < z_c$ \vspace*{4mm} \\
- \kappa \left(1 - \frac{1+z_c}{1+z} \rme^{-(|k-\half|+\half)/\xi}\right)
\left[Q^+_{k} \Theta_{k} + Q^-_{k} \bar{\Theta}_{k}\right] 
& $z > z_c$,
}
\label{TheoSreflc}
\end{eqnarray}
\normalsize
and the boundary layer part
\small
\begin{eqnarray}
\tilde{S}^{\mathrm{bl}}_{\infty}(k,l) 
& = \cases{0 & $z < z_c$ \vspace*{4mm} \\
 - \frac{z-z_c}{1+z_c} \frac{\kappa}{z} \Theta_{k+l-1} 
\left(\rme^{-l/\xi} Q^-_{k} \bar{\Theta}_{k}\Theta_{l} 
+ \rme^{-k/\xi} Q^-_{l} \Theta_{k} \bar{\Theta}_{l} \right)
& \\
+ \frac{\kappa}{z} \rme^{-l/\xi} 
\left[ \left(\frac{1+z}{1+z_c} - \rme^{-k/\xi}\right) 
Q^+_{k} \right. 
& \\
\left. + \left(1 - \rme^{-k/\xi}\right) 
Q^-_{k} \right] \Theta_{l-k} \Theta_{k}\Theta_{l}
& \\
+ \rme^{-k/\xi} 
\left[ \left(\frac{1+z}{1+z_c} - \rme^{-l/\xi}\right) 
Q^+_{l} \right. 
& \\
\left. + \left(1 - \rme^{-l/\xi}\right) 
Q^-_{l} \right] \Theta_{k-l} \Theta_{k}\Theta_{l}
& $z > z_c$.
}
\label{TheoSL0c} 
\end{eqnarray}
\normalsize
\end{theorem}

\begin{remark}
The picture is reverted compared to the behaviour near the blockage: 
Correlations are equal to the bulk values of Theorem 
\ref{Theo:bulkcorrelation} below the critical point and come arbitrarily 
close to these bulk values above the critical point as $\max{\{k,l\}}$ becomes 
large compared to the localization length $\xi$. The boundary layer 
contribution to the correlation function is restricted to sector 2 due to particle 
number conservation and the choice $\rho \leq 1/2$.
\end{remark}

To get insight about the critical point we consider hydrodynamic scaling
for the sublattice correlations defined for $1 \leq m,n \leq L/2$ by
\begin{eqnarray}
S^{++}_L(m,n) := S_L(2m,2n), \qquad & S^{+-}_L(m,n) := S_L(2m,2n-1) \\
S^{-+}_L(m,n) := S_L(2m-1,2n), \qquad & S^{--}_L(m,n) := S_L(2m-1,2n-1) .
\end{eqnarray}
This yields the sublattice decomposition
\begin{eqnarray}
S_L(k,l) & = 
S^{++}_L(k/2,l/2) Q_k^+ Q_l^+ 
+ S^{--}_L(\floor{k/2}+1,\floor{l/2}+1) Q_k^- Q_l^- \nonumber \\
& + S^{+-}_L(k/2,\floor{l/2}+1) Q_k^+ Q_l^- 
+ S^{-+}_L(\floor{k/2}+1,l/2) Q_k^- Q_l^+ 
\label{SLklsub}
\end{eqnarray}
of the density correlation function.

\begin{theorem}[Correlations on macroscopic scale]
\label{Theo:correlationscrit} 
Let for $u,v\in(-1/2,1/2)$ 
\begin{eqnarray}
\tilde{A}^{\mathrm{hc,\pm}}(u) 
& := \lim_{L\to\infty} A^{\mathrm{hc,\pm}}_L(L/4+\floor{uL/2}) \\
\tilde{A}^{\mathrm{refl,\pm}}(u) 
& := \lim_{L\to\infty} A^{\mathrm{refl,\pm}}_L(L/4+\floor{uL/2}) \\
\tilde{S}^{\mathrm{bl\pm\pm}}(u,v) 
& := \lim_{L\to\infty} 
\tilde{S}^{\mathrm{bl\pm\pm}}_L(L/4+\floor{uL/2},L/4+\floor{vL/2}) 
\end{eqnarray}
be the centered hardcore, reflective, and boundary layer contributions to the 
sublattice density correlation function under hydrodynamic scaling. 
With the scaling functions
\begin{eqnarray}
\tilde{a}^{\mathrm{hc}}(u) & := \frac{2u}{1+z_c}
\left(1 - \frac{2u}{1+z_c} \right) \\
\tilde{a}^{\mathrm{refl}}(u) & := 2 |u| \\
\tilde{s}^{\mathrm{bl}}(u,v) & := 2 u \left(1 - 2v\right)
\end{eqnarray}
the non-vanishing contributions are the hardcore correlations
\begin{eqnarray}
\tilde{A}^{\mathrm{hc,-}}(u) & = 
\cases{\kappa & $z < z_c$ \\
\kappa - \tilde{a}^{\mathrm{hc}}(u) \Theta(u) & $z = z_c$ \\
\kappa \bar{\Theta}(u) & $z > z_c$
}
\label{tAcrithc-} \\
\tilde{A}^{\mathrm{hc},+}(u) & = 
\cases{
0 & $z < z_c$ \\
\tilde{a}^{\mathrm{hc}}(u)  \Theta(u) & $z = z_c$ \\
\kappa \Theta(u) & $z > z_c$,
}
\label{tAcrithc+} 
\end{eqnarray}
the long-range reflective anticorrelations
\begin{eqnarray}
\tilde{A}^{\mathrm{refl,-}}(u) & = \cases{
0 & $z < z_c$ \\
- \kappa \tilde{a}^{\mathrm{refl}}(u) \bar{\Theta}(u) 
& $z = z_c$ \\
- \kappa \bar{\Theta}(u) & $z > z_c$,
}
\label{tAcritrefl-} \\ 
\tilde{A}^{\mathrm{refl,+}}(u) & = \cases{
0 & $z < z_c$ \\
- \kappa \tilde{a}^{\mathrm{refl}}(u) \Theta(u)
& $z = z_c$ \\
- \kappa \Theta(u) & $z > z_c$,
}
\label{tAcritrefl+} 
\end{eqnarray}
and the boundary layer correlations 
\begin{eqnarray}
\tilde{S}^{\mathrm{bl\pm\pm}}(u,v) 
& = \cases{
0 & $z<z_c$ \\
\frac{\kappa}{z_c} 
\tilde{s}^{\mathrm{bl}}(u,v) \Theta(v-u) \Theta(u) \Theta(v) \nonumber \\
+ \frac{\kappa}{z_c} \tilde{s}^{\mathrm{bl}}(v,u)
\Theta(u-v) \Theta(u) \Theta(v) & $z=z_c$ \\
0 & $z>z_c$ 
} \nonumber \\
\label{tScrit0}
\end{eqnarray}
\normalsize
independently of the sublattices.
\end{theorem}

\begin{remark}
Since the width $\xi$ of the boundary layer diverges as one approaches the 
critical point, correlations extend over the full lattice sector 2. As worked out 
already in Theorem \ref{Theo:bulkcorrelation}, the boundary layer contribution 
vanishes away from the critical point on hydrodynamic scale since the 
macroscopic width $\lim_{L\to\infty}\xi/L$ of the boundary layer is zero
for $z\neq z_c$.
\end{remark}

\subsubsection{Exact finite-size density correlation function}

The proofs of all four theorems \ref{Theo:bulkcorrelation} -
\ref{Theo:correlationscrit} are based on taking appropriate limits 
$L\to\infty$ of the exact finite-size expression of the density correlation 
function \eref{Skl} established in Proposition \ref{Prop:SLkl} below. 

To express the functional dependence of finite-size density correlation 
function on the parameters $p,z,L,k,l$ we introduce as the auxiliary constants
\begin{eqnarray}
\Gamma_L & := j_L (j_{L-2} - j_L) 
\label{GammaLdef} \\
\Delta_L & := 1 - \frac{j_L}{p} 
\label{DeltaLdef}
\end{eqnarray}
written out in explicit form in \eref{GammaL}, \eref{DeltaL},
and the auxiliary functions
\begin{eqnarray}
\Psi_L(m) & := H_L(m+1) - H_L(m) 
\label{PsiLdef} \\
\tilde{\Psi}_L(m) & := \tilde{H}_L(m-1) - \tilde{H}_L(m) 
\label{tPsiLdef} \\
F_L(m,n) 
& := \frac{H_L(m)}{p} \left[1-p - \frac{H_L(n)}{p}\right] 
\label{FLmndef} \\
\tilde{F}_L(m,n) 
& := \frac{\tilde{H}_L(m)}{p} \left[1-p - \frac{\tilde{H}_L(n)}{p}\right] 
\label{tFLmndef} 
\end{eqnarray}
given in explicit form in \eref{PsiL}, \eref{FLmn}.
The SMPM then yields the following exact expressions.

\begin{proposition}
\label{Prop:SLkl}
For $m,n\in \{1,\dots,L/2\}$ the density correlation functions for any 
system size $L=4K$, $K \in \N$ is given by
\begin{eqnarray}
S_{L}(m,n) & = \Gamma_L Q^-_{m} Q^-_{n} 
+ \kappa_L Q^-_{m} \delta_{m,n} 
\label{SL11} \nonumber \\[2mm]
S_{L}(m,L+1-n) 
& = \Gamma_L Q^-_{m} Q^+_{n} - H_L(m) Q^-_{m} \delta_{m,n} \nonumber \\
& + \Delta_L H_L(n) Q^-_{m} + \Psi_L(n) Q^-_{m} \Theta_{m-n}
\label{SL12} \nonumber \\[2mm]
S_{L}(L+1-m,n) 
& = \Gamma_L Q^-_{n} Q^+_{m} - H_L(m) Q^-_{m} \delta_{m,n} \nonumber \\
& + \Delta_L H_L(m) Q^-_{n} + \Psi_L(m) Q^-_{n} \Theta_{n-m}
\label{SL21} \nonumber \\[2mm]
S_{L}(L+1-m,L+1-n)
& = \Gamma_L Q^+_{m} Q^+_{n} + \kappa_{L} Q^+_{m} \delta_{m,n} 
\nonumber \\
& + F_L(m,m) \delta_{m,n} + H_L(m) (Q^-_{m} - Q^+_{m}) \delta_{m,n} 
\nonumber \\
& + F_L(m,n) \Theta_{m-n} + F_L(n,m) \Theta_{n-m} \nonumber \\
& + \Psi_L(n) Q^+_{m} \Theta_{m-n} + \Psi_L(m) Q^+_{n} \Theta_{n-m} 
\nonumber \\ 
& + H_L(n) Q^+_{m} \Delta_L + \Delta_L H_L(m) Q^+_{n} .
\label{SL22} 
\end{eqnarray}
\end{proposition}

\proof We prove the proposition with a case-by-case computation of the
sublattice correlation functions using the SMPM \eref{MPA} and the properties 
\eref{Aqr1} - \eref{Lem1d} of the matrix algebra. In particular, we note that
from \eref{Lem1b} one gets
\begin{eqnarray}
\frac{p z}{Y_{\frac{L}{2}}} \Tr (D A^{n-1} A_2 A^{\frac{L}{2}-n}) 
= H_L(n)  , \qquad 1 \leq n \leq \frac{L}{2},
\label{WA2V} \\
\frac{z^2}{Y_{\frac{L}{2}}} \Tr (D A^{n-1} A_2 A^{m-n-1} A_2 A^{\frac{L}{2}-m})
= \frac{H_L(m)}{z_c} , \qquad 1 \leq n < m \leq \frac{L}{2}.
\label{WA2A2V}
\end{eqnarray} 
We also note that for $m,n\in\Z$
\begin{eqnarray}
\delta_{m,|n|} = \delta_{m,n} \Theta_{n} + \delta_{m,-n} \bar{\Theta}_{n}, \\
\label{nsym}
|m-1/2| + 1/2 = m \Theta_{m} + (|m|+1) \bar{\Theta}_{m}.
\label{msym}
\end{eqnarray}
In the following sublattice computations we assume throughout 
$k,l\in\{1,\dots,L/4\}$.

\paragraph{\underline{Odd-odd correlations}}

For the joint expectations one finds
\begin{eqnarray}
\exval{\eta_{2k-1} \eta_{2l-1}}_L 
& = \cases{j_{L-2} j_L & $k\neq l$ \\
j_L & $k = l$} 
\nonumber \\[2mm]
\exval{\eta_{2k-1} \eta_{L+1-2l}}_L 
& = \cases{j_{L-2} j_L + H_L(2l) & $k \leq l$\\
j_{L-2} j_L + H_L(2l+1) & $k \geq l+1$} 
\nonumber  \\[2mm]
\exval{\eta_{L+1-2k} \eta_{2l-1}}_L 
& = \cases{j_{L-2} j_L + H_L(2k) & $l \leq k$\\
j_{L-2} j_L + H_L(2k+1) & $l \geq k+1$} 
\nonumber \\[2mm]
\exval{\eta_{L+1-2k} \eta_{L+1-2l}}_L 
& = \cases{
j_{L-2}j_{L} + \frac{1}{p} H_L(2l) + H_L(2k+1)  & 
$k < l$\\
j_{L} + \frac{1}{p} H_L(2l)  & $k = l$\\
j_{L-2}j_{L} + \frac{1}{p} H_L(2k) + H_L(2l+1)
& $k > l$ .
} 
\nonumber
\end{eqnarray}
With the exact expression \eref{rhoLn} for the density profile it follows that
\begin{eqnarray}
S_L(2k-1,2l-1)
& = \cases{
\Gamma_L 
& $k\neq l$ \\
\Gamma_L + \kappa_L  
& $k = l$
} 
\nonumber \\[2mm]
S_L(2k-1,L+1-2l)
& = \cases{
\Gamma_L  + \Delta_L H_L(2l)
& $k \leq l$ \\
\Gamma_L + \Delta_L H_L(2l) + \Psi_L(2l) 
& $k > l$} 
\nonumber \\[2mm]
S_L(L+1-2k,2l-1)
& = \cases{
\Gamma_L + \Delta_L H_L(2k) + \Psi_L(2k) 
& $k < l$ \\
\Gamma_L  + \Delta_L H_L(2k) 
& $k \geq l$ 
} 
\nonumber \\[2mm]
S_L(L+1-2k,L+1-2l)
& = \cases{
\Gamma_L + F_L(2l,2k) + \Psi_L(2k) & \\
+ \Delta_L H_L(2k) + \Delta_L H_L(2l) 
& $k < l$ \\
\Gamma_{L} + \kappa_{L} + F_L(2k,2k) & \\
+ \left(2\Delta_{L} - 1\right) H_L(2l)
& $k = l$ \\
\Gamma_L + F_L(2k,2l) + \Psi_L(2l) & \\
+ \Delta_L H_L(2k) + \Delta_L H_L(2l)
& $k > l$ .
} 
\nonumber 
\end{eqnarray}

\paragraph{\underline{Odd-even correlations:}}
From (\ref{smpm1b}) in Proposition \ref{Prop:smpm} one obtains
\begin{equation}
\exval{\eta_{L+1-2k} \eta_{L+2-2l}}_L = \exval{\eta_{L+2-2l}}_L .
\end{equation}
With this and (\ref{etaL+2-2k}) the SMPM yields 
\begin{eqnarray}
\exval{\eta_{2k-1} \eta_{2l}}_L 
& = 0 
\nonumber \\[2mm]
\exval{\eta_{2k-1} \eta_{L+2-2l}}_L 
& = \cases{
H_L(2l-1) & $k < l$\\
0 & $k = l$\\
H_L(2l) & $k > l$
}  
\nonumber \\[2mm]
\exval{\eta_{L+1-2k} \eta_{2l}}_L 
& = 0 
\nonumber \\[2mm]
\exval{\eta_{L+1-2k} \eta_{L+2-2l}}_L 
& = \cases{
\frac{1}{p} H_L(2l-1) & $k < l$ \\
H_L(2l) + \frac{H_L(2k) }{z_c}  & $k \geq l$ .
}
\nonumber 
\end{eqnarray}
It follows that
\begin{eqnarray}
S_L(2k-1,2l)
& = 0 
\nonumber \\[2mm]
S_L(2k-1,L+2-2l)
& = \cases{
\Delta_L H_L(2l-1) & $k < l$ \\
\left(\Delta_L - 1\right) H_L(2l-1) & $k = l$ \\
\Delta_L H_L(2l-1) + \Psi_L(2l-1) & $k > l$
}  
\nonumber \\[2mm]
S_L(L+1-2k,2l)
& = 0 
\nonumber \\[2mm]
S_L(L+1-2k,L+2-2l)
& = \cases{
F_L(2l-1,2k) & \\
+ \Delta_L H_L(2l-1) & $k < l$ \\
F_L(2k,2l-1) + \Psi_L(2l-1) & \\
+ \Delta_L H_L(2l-1) & $k \geq l$.
} 
\nonumber 
\end{eqnarray}
Similarly, by the symmetry \eref{Slk} one has
\small
\begin{eqnarray}
S_L(2k,2l-1)
& = 0 
\nonumber \\[2mm]
S_L(2k,L+1-2l)
& = 0 
\nonumber \\[2mm]
S_L(L+2-2k,2l-1)
& = \cases{
\Delta_L H_L(2k-1) + \Psi_L(2k-1) & $k < l$ \\
\left(\Delta_L - 1\right) H_L(2k-1) & $k = l$ \\
\Delta_L H_L(2k-1)
& $k > l$ 
} 
\nonumber \\[2mm]
S_L(L+2-2k,L+1-2l)
& = \cases{
F_L(2l,2k-1) + \Psi_L(2k-1) & \\
+ \Delta_L H_L(2k-1) & $k \leq l$\\
F_L(2k-1,2l) + \Delta_L H_L(2k-1) & $k > l$ .
} 
\nonumber 
\end{eqnarray}
\normalsize

\paragraph{\underline{Even-even correlations:}}

From the projection property (\ref{smpm1a}) and \eref{WA2A2V} one finds
\begin{eqnarray}
\exval{\eta_{2k} \eta_{2l}}_L & = 0 \nonumber \\
\exval{\eta_{2k} \eta_{L+2-2l}}_L & = 0 \nonumber \\
\exval{\eta_{L+2-2k} \eta_{2l}}_L & = 0 \nonumber \\
\exval{\eta_{L+2-2k} \eta_{L+2-2l}}_L 
& = \cases{
\frac{H_L(2l-1)}{z_c} & $k < l$\\
\frac{1}{p} H_L(2k-1) & $k = l$\\
\frac{H_L(2k-1)}{z_c} & $k > l$ .
} 
\nonumber 
\end{eqnarray}
Therefore
\begin{eqnarray}
S_L(2k,2l) & = 0 \nonumber \\[2mm]
S_L(2k,L+2-2l) & = 0 \nonumber \\[2mm]
S_L(L+2-2k,2l) & = 0 \nonumber \\[2mm]
S_L(L+2-2k,L+2-2l)
& = \cases{
F_L(2l-1,2k-1) & $k < l$ \\
F_L(2k-1,2k-1) & \\
+ H_L(2k-1) & $k = l$ \\
F_L(2k-1,2l-1) & $k > l$ .
}
\nonumber 
\end{eqnarray}

Adding up the parts of the correlation function according to the 
sublattice decomposition \eref{SLklsub} proves the proposition. 
\qed

\subsubsection{Proof of the theorems \ref{Theo:bulkcorrelation}, 
\ref{Theo:correlationsd}, \ref{Theo:correlationsc}, \ref{Theo:correlationscrit}}

It is convenient to introduce for $n\in\Z$ the sector indicator functions
\begin{eqnarray}
\Theta^{(1)}_{L,n} := \sum_{k=1}^{L/2} \delta_{k,n}, \qquad
\Theta^{(0)}_{L,n} := \Theta^{(1)}_{L,n+L/2}, \qquad
\Theta^{(2)}_{L,n} := \Theta^{(1)}_{L,n-L/2}
\label{indidef}
\end{eqnarray}
From Proposition \ref{Prop:SLkl} one finds that one has for $k,l\in\T_L$
\begin{eqnarray}
S_{L}(k,l) & = \left[\Gamma_L Q^-_{k} Q^-_{l} 
+ \kappa_L  Q^-_{k} \delta_{k,l}\right] \Theta^{(1)}_{L,k}\Theta^{(1)}_{L,l}
\nonumber \\[2mm]
& + \left[\Gamma_L Q^-_{k} Q^-_{l} 
- H_L(k) Q^-_{k} \delta_{k+l,L+1} \right. \nonumber \\
& + \Delta_L H_L(L+1-l) Q^-_{k} \nonumber \\
& \left. + \Psi_L(L+1-l) Q^-_{k} \Theta(k+l-L-1)\right]
\Theta^{(1)}_{L,k}\Theta^{(2)}_{L,l}
\nonumber \\[2mm]
& + \left[\Gamma_L  Q^-_{k} Q^-_{l}
- H_L(L+1-k) Q^+_{k} \delta_{k+l,L+1} \right. \nonumber \\
& + \Delta_L H_L(L+1-k) Q^-_{l} \nonumber \\
& \left. + \Psi_L(L+1-k) Q^-_{l} \Theta(k+l-L-1)\right]
\Theta^{(2)}_{L,k}\Theta^{(1)}_{L,l}
\nonumber \\[2mm]
& + \left[\Gamma_L Q^-_{k} Q^-_{l} + \kappa_{L} Q^-_{k} \delta_{k,l} 
\right. \nonumber \\
& + F_L(L+1-k,L+1-k) \delta_{k,l} 
+ H_L(L+1-k) (Q^+_{k} - Q^-_{k}) \delta_{k,l} \nonumber \\
& + \Psi_L(L+1-l) Q^-_{k}\Theta_{l-k} 
+ \Psi_L(L+1-k) Q^-_{l} \Theta_{k-l} \nonumber \\ 
& + \Delta_L H_L(L+1-l) Q^-_{k} 
+ \Delta_L H_L(L+1-k) Q^-_{l} \nonumber \\
& + F_L(L+1-k,L+1-l)\Theta_{l-k} \nonumber \\
& \left. + F_L(L+1-l,L+1-k) \Theta_{k-l} \right] \Theta^{(2)}_{L,k}\Theta^{(2)}_{L,l}.
\label{SLkl} 
\end{eqnarray}
One reads off
\begin{eqnarray}
A^{\mathrm{hc}}_{L}(k)  & = (\kappa_L+\Gamma_L) Q^-_{k}  
+ H_L(L+1-k) (Q^+_{k} + (2\Delta_L-1) Q^-_{k}) \Theta^{(2)}_{L,k} 
\nonumber \\
& + F_L(L+1-k,L+1-k)  \Theta^{(2)}_{L,k}
\label{AhcLk} \\
A^{\mathrm{refl}}_{L}(k) & = (\Delta_L-1) \left[H_L(k) Q^-_k \Theta^{(1)}_{L,k} 
+ H_L(L+1-k) Q^+_k \Theta^{(2)}_{L,k}\right]
\label{AreflLk} \\
S^{\mathrm{bl}}_{L}(k,l) & = \Gamma_L Q^-_{k} Q^-_{l} (1-\delta_{k,l}) 
\nonumber \\
& + \Delta_L H_L(L+1-l) Q^-_{k} \Theta^{(1)}_{L,k}\Theta^{(2)}_{L,l} \nonumber \\
& + \Psi_L(L+1-l) Q^-_{k} \Theta(k+l-L-1) \Theta^{(1)}_{L,k}\Theta^{(2)}_{L,l}
\nonumber \\
& + \Delta_L H_L(L+1-k) Q^-_{l} \Theta^{(2)}_{L,k}\Theta^{(1)}_{L,l} \nonumber \\
& + \Psi_L(L+1-k) Q^-_{l} \Theta(k+l-L-1) \Theta^{(2)}_{L,k}\Theta^{(1)}_{L,l}
\nonumber \\
& + \left[ \Psi_L(L+1-l) Q^-_{k}\Theta_{l-k} 
+ \Psi_L(L+1-k) Q^-_{l} \Theta_{k-l} \right. \nonumber \\ 
& + \Delta_L \left( H_L(L+1-l) Q^-_{k} 
+ H_L(L+1-k) Q^-_{l}\right) (1-\delta_{k,l})\nonumber \\
& + F_L(L+1-k,L+1-l)\Theta_{l-k} \nonumber \\
& \left. + F_L(L+1-l,L+1-k) \Theta_{k-l} \right] \Theta^{(2)}_{L,k}\Theta^{(2)}_{L,l}.
\label{SblLkl} 
\end{eqnarray}
which yields (i) for the hard core part
\small
\begin{eqnarray}
A^{\mathrm{hc}}_{L}(k) \Theta^{(1)}_{L,k} & = (\kappa_L+\Gamma_L)  Q^-_{k}  \Theta^{(1)}_{L,k}
\label{AhcLkp} \\
A^{\mathrm{hc}}_{L}(L-|k|) \Theta^{(0)}_{L,k} & = \left[(\kappa_L+\Gamma_L)  Q^-_{k} + F_L(|k|+1,|k|+1) \right] \Theta^{(0)}_{L,k} 
\nonumber \\
& + H_L(|k|+1) (Q^+_{k} + (2\Delta_L-1) Q^-_{k}) \Theta^{(0)}_{L,k}
\label{AhcLkm} \\
A^{\mathrm{hc}}_{L}(L/2+k) & = (\kappa_L+\Gamma_L) Q^-_{k} 
 + \tilde{F}_L(|k|,|k|) \Theta^{(1)}_{L,k} \nonumber \\
& + \tilde{H}_L(|k|) (Q^+_{k} + (2\Delta_L-1)  Q^-_{k}) \Theta^{(1)}_{L,k}, 
\label{tAhcLk} 
\end{eqnarray}
\normalsize
(ii) for the reflective part
\small
\begin{eqnarray}
A^{\mathrm{refl}}_{L}(k) \Theta^{(1)}_{L,k} & = (\Delta_L-1) H_L(|k|) Q^-_k \Theta^{(1)}_{L,k} 
\label{AreflLkp} \\
A^{\mathrm{refl}}_{L}(L-|k|)  \Theta^{(2)}_{L,k} & = (\Delta_L-1)
H_L(|k|+1) Q^+_k \Theta^{(0)}_{L,k}
\label{AreflLkm} \\
A^{\mathrm{refl}}_{L}(L/2+k) & = (\Delta_L-1) \tilde{H}_L(|k|+1) Q^-_k \Theta^{(0)}_{L,k} \nonumber \\
& + (\Delta_L-1) \tilde{H}_L(|k|) Q^+_k \Theta^{(1)}_{L,k} ,
\label{tAreflLk} 
\end{eqnarray}
\normalsize
and (iii) for the boundary layer part
\small
\begin{eqnarray}
S^{\mathrm{bl}}_{L}(k,l) \Theta^{(1)}_{L,k}\Theta^{(1)}_{L,l} 
& = \Gamma_L Q^-_{k} Q^-_{l} \Theta^{(1)}_{L,k}\Theta^{(1)}_{L,l} (1-\delta_{k,l})
\label{SLklblpp} \\
S^{\mathrm{bl}}_{L}(k,L-|l|) \Theta^{(1)}_{L,k}\Theta^{(0)}_{L,l} 
& = \Delta_L H_L(|l|+1) Q^-_{k} \Theta^{(1)}_{L,k}\Theta^{(0)}_{L,l} \nonumber \\
& + \Psi_L(|l|+1) Q^-_{k} \Theta(k+l-1) \Theta^{(1)}_{L,k}\Theta^{(0)}_{L,l}
\nonumber \\
& + \Gamma_L Q^-_{k} Q^-_{l}  \Theta^{(1)}_{L,k}\Theta^{(0)}_{L,l}
\label{SLklblpm} \\
S^{\mathrm{bl}}_{L}(L-|k|,l) \Theta^{(0)}_{L,k} \Theta^{(1)}_{L,l} 
& =  H_L(|k|+1) Q^-_{l} \Theta^{(0)}_{L,k} \Theta^{(1)}_{L,l} \nonumber \\
& + \Psi_L(|k|+1) Q^-_{l} \Theta(k+l-1) \Theta^{(0)}_{L,k} \Theta^{(1)}_{L,l}
\nonumber \\
& + \Gamma_L Q^-_{k} Q^-_{l} \Theta^{(0)}_{L,k} \Theta^{(1)}_{L,l} 
\label{SLklblmp} \\
S^{\mathrm{bl}}_{L}(L-|k|,L-|l|) \Theta^{(0)}_{L,k}\Theta^{(0)}_{L,l} 
& = \left[ \Delta_L H_L(|l|+1) Q^-_{k} 
+ \Delta_L H_L(|k|+1) Q^-_{l}  \right. \nonumber \\
& + \Psi_L(|l|+1) Q^-_{k}\Theta_{l-k} 
+ \Psi_L(|k|+1) Q^-_{l} \Theta_{k-l} \nonumber \\ 
& + F_L(|k|+1,|l|+1)\Theta_{l-k} \nonumber \\
& \left. + F_L(|l|+1,|k|+1) \Theta_{k-l} \Theta^{(0)}_{L,k}\Theta^{(0)}_{L,l}\right] 
\nonumber \\
& + \Gamma_L Q^-_{k} Q^-_{l} (1-\delta_{k,l}) 
\Theta^{(0)}_{L,k}\Theta^{(0)}_{L,l}  .
\label{SLklblmm} \\
S^{\mathrm{bl}}_{L}(L/2+k,L/2+l) & = \Delta_L \tilde{H}_L(l) Q^-_{k}  (1-\delta_{k+l,1})
\Theta^{(0)}_{L,k}\Theta^{(1)}_{L,l} \nonumber \\
& + \Delta_L \tilde{H}_L(k) Q^-_{l} (1-\delta_{k+l,1}) \Theta^{(1)}_{L,k}\Theta^{(0)}_{L,l} \nonumber \\
& + \tilde{\Psi}_L(l) Q^-_{k} \Theta(k+l-1) \Theta^{(0)}_{L,k}\Theta^{(1)}_{L,l}
\nonumber \\
& + \tilde{\Psi}_L(k) Q^-_{l} \Theta(k+l-1) \Theta^{(1)}_{L,k}\Theta^{(0)}_{L,l}
\nonumber \\
& + \left[ \tilde{\Psi}_L(l) Q^-_{k}\Theta_{l-k} 
+ \tilde{\Psi}_L(k) Q^-_{l} \Theta_{k-l} \right. \nonumber \\ 
& + \Delta_L \left(\tilde{H}_L(l) Q^-_{k} 
+ \tilde{H}_L(k) Q^-_{l} \right) (1-\delta_{k,l}) \nonumber \\
& \left. + \tilde{F}_L(k,l)\Theta_{l-k} + \tilde{F}_L(l,k) \Theta_{k-l} \right] \Theta^{(1)}_{L,k}\Theta^{(1)}_{L,l}
\nonumber \\
& + \Gamma_L Q^-_{k} Q^-_{l} (1-\delta_{k,l}) 
\label{tSLklbl}
\end{eqnarray}
\normalsize
The offcritical thermodynamic limits relevant for theorems 
\ref{Theo:bulkcorrelation} - \ref{Theo:correlationsc}
are readily computed from these exact finite-size expressions by
using the asymptotic values \eref{kappa} and those derived in \ref{App:lim}.

At the critical point where $\kappa = p(1-p)$ and $(1-p)^2 = \kappa/z_c$
one gets from Proposition \ref{Prop:SLkl} and the large-$L$ results 
derived in \ref{App:lim} to leading order in $1/L$
\begin{eqnarray}
S_{L}(L/2+k,L/2+l) 
& =  \kappa Q^-_{k} \Theta^{(0)}_{L,k} \delta_{k,l}  \nonumber \\
& + (1-p) \left(1-\frac{2k}{L}\right) 
 \left(p - (1-p)\frac{2k}{L}\right)  
Q^-_{k} \Theta^{(1)}_{L,k} \delta_{k,l}  \nonumber \\
& + (1-p) \frac{2k}{L} \left(1 - (1-p) \frac{2k}{L}\right)  Q^+_{k} \Theta^{(1)}_{L,k} \delta_{k,l}  \nonumber \\
& - \kappa \frac{2|k|}{L} \left[Q^-_{k} \bar{\Theta}_{L,1}(-k) + Q^+_{k} \Theta^{(1)}_{L,k} \right] \delta_{k+l,1} \nonumber \\
& + \frac{\kappa}{z_c} \frac{2k}{L} \left(1-\frac{2l}{L}\right)\Theta_{l-k} \Theta^{(1)}_{L,k} \Theta^{(1)}_{L,l} \nonumber \\
& + \frac{\kappa}{z_c} \frac{2l}{L} \left(1-\frac{2k}{L}\right) \Theta_{k-l} \Theta^{(1)}_{L,k} \Theta^{(1)}_{L,l} .
\label{SL22crit} 
\end{eqnarray}

One reads off
\begin{eqnarray}
\tilde{A}^{\mathrm{hc}}_L(k) 
& = \kappa Q^-_{k} \Theta^{(0)}_{L,k} \nonumber \\
& + (1-p) \left(1-\frac{2k}{L}\right) 
 \left(p - (1-p)\frac{2k}{L}\right)  
Q^-_{k} \Theta^{(1)}_{L,k}  \nonumber \\
& + (1-p) \frac{2k}{L} \left(1 - (1-p) \frac{2k}{L}\right)  Q^+_{k} \Theta^{(1)}_{L,k}   \\
\tilde{A}^{\mathrm{refl}}_L(k) & = - \kappa \frac{2|k|}{L} \left[Q^-_{k} \Theta^{(0)}_{L,k} + Q^+_{k} \Theta^{(1)}_{L,k} \right] \\
\tilde{S}^{\mathrm{bl}}_L(k,l) & = \frac{\kappa}{z_c} \frac{2k}{L} 
\left(1-\frac{2l}{L}\right)\Theta_{l-k} \Theta^{(1)}_{L,k} \Theta^{(1)}_{L,l} 
\nonumber \\
& + \frac{\kappa}{z_c} \frac{2l}{L} \left(1-\frac{2k}{L}\right) \Theta_{k-l} 
\Theta^{(1)}_{L,k} \Theta^{(1)}_{L,l}.
\end{eqnarray}

Projecting on the sublattices and taking the scaling limit with $k = \floor{uL}$
and $l = \floor{vL}$ where $u,v \in (-1/2,1/2]$ yields 
\eref{tAcrithc-} - \eref{tScrit0}. \qed

\subsection{Static structure function}

The static structure function \eref{SLrdef} has recently turned 
out to be of interest in the context of hydrodynamic scaling \cite{Kare19}.
Here we use it to shed light on the behaviour of the variance 
established in Theorem \ref{Theo:rhovariance}. 

\subsubsection{Synopsis}

It is shown that the reflective anticorrelations are responsible for 
the vanishing compressibility in the phase-separated regime $z>z_c$. 
Above the critical point, these anticorrelations exactly cancel the hard core 
contribution. Nevertheless, locally the offcritical static structure function 
reduces in the thermodynamic limit to its hard-core contribution, as if 
(erroneously) correlations produced by the blockage were irrelevant.
At the critical point the static structure function has a non-trivial scaling form
due to the macroscopic size of the critical boundary layers which also leads
to the divergent critical compressibility. In the free flow phase below the 
critical point the compressibility is fully determined by the hard core
part of the static structure function. Both at and off criticality, the static 
structure function has no sublattice dependence in the limit $L\to\infty$.

\subsubsection{Main results}

We recall that $S_L(r) = S_L(-r)$ so that is sufficient to consider
$0 \leq r \leq L/2$. Guided by the results on the two-point density correlation 
function we decompose the static structure function \eref{SLrdef} as
\begin{equation}
S_L(r) = S^{\mathrm{hc}}_L(r) + S^{\mathrm{refl}}_L(r)
+ S^{\mathrm{bl}}_L(r) 
\end{equation} 
and define the limits
\begin{eqnarray}
S_\infty(r) := \lim_{L\to\infty} S_L(r), \qquad 
S_\infty^{\mathrm{hc}}(r) := \lim_{L\to\infty} S^{\mathrm{hc}}_L(r) \\
S_\infty^{\mathrm{refl}}(r) := \lim_{L\to\infty} S^{\mathrm{refl}}_L(r), \qquad 
S_\infty^{\mathrm{bl}}(r) := \lim_{L\to\infty} S^{\mathrm{bl}}_L(r).
\label{Srdef}
\end{eqnarray} 

\begin{theorem}[Offcritical static structure function]
\label{Theo:Ossf}
For $z \neq z_c$ the reflective contribution $S_\infty^{\mathrm{refl}}(r)$
and the boundary layer contribution $S_\infty^{\mathrm{bl}}(r)$ 
to static structure function $S_\infty(r)$ vanish in the thermodynamic limit
and one has
\begin{equation}
S_\infty(r) = S_\infty^{\mathrm{hc}}(r) 
= \frac{\kappa}{2} \delta_{r,0}, \qquad r \in \Z
\label{Sr}
\end{equation}
with amplitude $\kappa$ given in \eref{kappa}.
\end{theorem}

\begin{remark}
As shown in \ref{App:lim}, the contributions of the reflective long-range 
anticorrelation \eref{AreflLk} and of the boundary layer part \eref{SblLkl} 
to the dynamical structure function \eref{SLrdef} in the offcritical regime are 
of order $1/L$ for any fixed $r$ and hence vanish in the thermodynamic limit.
\end{remark}

Next we consider hydrodynamic scaling and define for $|u| \in (0,1/2)$
the limits
\begin{eqnarray}
S^+(u) := \lim_{L\to\infty} S_L(2[uL/2]), \qquad 
S^-(u) := \lim_{L\to\infty} S_L(2[uL/2]-1)
\label{Spmudef}
\end{eqnarray} 
and analogously $S^{\mathrm{hc}\pm}(u)$, $S^{\mathrm{refl}\pm}(u)$,
and $S^{\mathrm{bl}\pm}(u)$. 

\begin{theorem}[Critical static structure function]
\label{Theo:Cssf}
At the critical point $z=z_c$ the hard-core contribution 
$S^{\mathrm{hc}\pm}(u)$ and the reflective contribution 
$S^{\mathrm{refl}\pm}(u)$ to the static structure function $S^\pm(u)$ 
vanish under hydrodynamic scaling for any macroscopic distance $u\neq 0$ 
and $|u| \in (0,1/2)$. One has
\begin{eqnarray}
S^{\pm}(u) = S^{\mathrm{bl}\pm}(u) = \frac{1}{12} (1-p)^2 (1-2|u|)^3 
\label{Sblu}
\end{eqnarray}
independently of the sublattice.
\end{theorem}

\begin{remark}
The hard core contribution $S^{\mathrm{hc}\pm}(u)$ vanishes by definition
for $u\neq 0$ while the reflective long-range contribution 
$S^{\mathrm{refl}\pm}(u)$ vanishes since in a finite system its contribution 
to the static structure function is, like in the off-critical case, of order $1/L$.
The critical fluctuations are dominated by the contribution from the
boundary layer.
\end{remark}

By definition, the particle variance \eref{Def:rhovariance} is given in terms
of the dynamical structure by $C_L = \sum_{r=-L/2+1}^{L/2} S_L(r)$.
Hence the decomposition $C_L = C^{\mathrm{hc}}_L + C^{\mathrm{refl}}_L
+ C^{\mathrm{bl}}_L$ and the corresponding limit
\begin{equation}
C = C^{\mathrm{hc}} + C^{\mathrm{refl}}
+ C^{\mathrm{bl}}
\end{equation}
provides insight into the origin of the fluctuations of the total particle number
in the grandcanonical ensemble.

\begin{theorem}[Particle number fluctuations] 
\label{Theo:Pnf}
The compressibility 
\eref{Def:rhovariance} has hard core, reflective, and boundary layer 
contributions given by
\begin{eqnarray}
C^{\mathrm{hc}} & =  \cases{
C & $z<z_c$ \\
\frac{p(1-p)}{2} + \frac{1}{12} (1-p)^2 & $z=z_c$ \\
\frac{p(1-p)}{2} & $z>z_c$ 
}
\label{Chc} \\
C^{\mathrm{refl}} & = \cases{
0 & $z<z_c$ \\
- \frac{p(1-p)}{4} & $z=z_c$ \\
- \frac{p(1-p)}{2} & $z>z_c$ 
}
\label{Crefl} \\
C^{\mathrm{bl}} & = \cases{
0 & $z<z_c$ \\
\infty & $z=z_c$ \\
0 & $z>z_c$ .
}
\label{Cbl}
\end{eqnarray}
At the critical point, the scaled variance of the particle number has the limiting
behaviour
\begin{equation}
\lim_{L\to\infty} \frac{1}{L} C^{\mathrm{bl}} = \frac{(1-p)^2}{48}.
\label{sCbl}
\end{equation}
\end{theorem}

\begin{remark}
The limit \eref{sCbl} of the scaled variance which arises from the the boundary 
layer contribution alone is equal to the scaled total variance \eref{rhovariance2} 
established in Theorem \ref{Theo:rhovariance},  thus
showing that the origin of the divergence of compressibility $C$ comes from
the unbounded fluctuations of the domain wall position in the thermodynamic 
limit.
\end{remark}

\subsubsection{Proofs}

To deal with the sector dependence of the density correlation function we 
split the sum \eref{SLrdef} defining static structure function as
\begin{eqnarray}
S_L(r) & = \sum_{k=1}^{L/2-r} S_L(k,k+r) 
+ \sum_{k=L/2-r+1}^{L/2} S_L(k,k+r) \nonumber \\
& + \sum_{k=L/2+1}^{L-r} S_L(k,k+r) 
+ \sum_{k=L-r+1}^{L} S_L(k,k+r-L) .
\label{SLrdec}
\end{eqnarray}
We also define for $0 \leq n \leq L/2$ further auxiliary functions
\begin{eqnarray}
G_L(n) & := \frac{2}{L} \sum_{m=1}^{L/2-n} F_L(m-n,m)
\label{GLndef} \\
\Phi^\pm_L(n) & := \frac{1}{L} \sum_{m=1}^{n} \Psi(m) Q^\pm_m . 
\label{PhiLndef}
\end{eqnarray}
Thermodynamic limits below are computed using the results of \ref{App:lim}.

\paragraph{\underline{Proof of Theorems \ref{Theo:Ossf} and 
\ref{Theo:Cssf}:}}
The hard core part trivially vanishes for $r = 0$ and with
\begin{equation}
B^{\mathrm{hc}}_L := \frac{1}{L} \sum_{k=1}^L A^{\mathrm{hc}}_L(k)
\end{equation}
one gets the exact finite-size result
\begin{equation}
S^{\mathrm{hc}}_L(r) :=  B^{\mathrm{hc}}_L \delta_{r,0}.
\end{equation}
From \eref{AhcLk} one finds
\begin{eqnarray}
B^{\mathrm{hc}}_{L} 
& = \frac{1}{2} \left[\Gamma_L + \kappa_L + H^-_L +  (2 \Delta_L-1) H^+_L  + G_L(0) \right].
\end{eqnarray}
Taking the limit $L\to\infty$ yields
\begin{eqnarray}
B^{\mathrm{hc}} =  \cases{
\frac{\kappa}{2} & $z<z_c$ \\
\frac{\kappa}{2} + \frac{1}{12} (1-p)^2 & $z=z_c$ \\
\frac{\kappa}{2} & $z>z_c$ 
}
\label{Bhc}
\end{eqnarray}
which proves the second equality of \eref{Sr} in Theorem \ref{Theo:Ossf} 
and \eref{Chc} in Theorem \ref{Theo:Pnf}.

For the reflective part one has trivially $S^{\mathrm{refl}}_L(2n) = 0$.
Proposition \ref{Prop:SLkl} yields for odd distances $r=2n-1$ after a brief 
calculation
\begin{eqnarray}
S^{\mathrm{refl}}_L(2n-1) 
& = \frac{1}{L} \left[A^{\mathrm{refl}}_L(L/2+1-n) + A^{\mathrm{refl}}_L(L+1-n)\right].
\end{eqnarray}
With \eref{AreflLk} we get
\begin{equation}
S^{\mathrm{refl}}_L(r) =  \frac{1}{L} (\Delta_L-1) 
\left(\tilde{H}_L(\floor{|r/2|}) Q^+_{\floor{|r/2|}} 
+ H_L(\floor{|r/2|}) Q^-_{\floor{|r/2|}} \right) Q^-_r.
\label{SLrrefl}
\end{equation}
and therefore in the limit $L\to\infty$ one has, independently of the sublattice,
\begin{equation}
S^{\mathrm{refl}}(r) = 0, \qquad S^{\mathrm{refl\pm}}(u) = 0
\end{equation}
as stated in Theorem \ref{Theo:Ossf} and in Theorem \ref{Theo:Cssf}.

Next we compute the sublattice parts of the boundary layer distribution.
For even distance $r=2n > 0$ one gets from Proposition \ref{Prop:SLkl}
\begin{eqnarray}
S^{\mathrm{bl}}_{L}(k,k+2n) & = \Gamma_L Q^-_{k}
\nonumber \\
& + \Delta_L H_L(L+1-k-2n) Q^-_{k} \Theta^{(1)}_{L,k}\Theta^{(2)}_{L,k+2n} \nonumber \\
& + \Psi_L(L+1-k-2n) Q^-_{k} \Theta(2k+2n-L-1) \Theta^{(1)}_{L,k}\Theta^{(2)}_{L,k+2n}
\nonumber \\
& + \Delta_L H_L(L+1-k) Q^-_{k} \Theta^{(2)}_{L,k}\Theta^{(1)}_{L,k+2n} \nonumber \\
& + \Psi_L(L+1-k) Q^-_{k} \Theta(2k+2n-L-1) \Theta^{(2)}_{L,k}\Theta^{(1)}_{L,k+2n}
\nonumber \\
& + \left[ \Psi_L(L+1-k-2n) Q^-_{k}\Theta_{2n} \right. \nonumber \\ 
& + \left(\Delta_L H_L(L+1-k-2n) Q^-_{k} 
+ \Delta_L H_L(L+1-k) Q^-_{k}\right) \nonumber \\
& \left. + F_L(L+1-k,L+1-k-2n)\Theta_{2n} \right] 
\Theta^{(2)}_{L,k}\Theta^{(2)}_{L,k+2n}.
\end{eqnarray}
Evaluating the individual sums \eref{SLrdec} and adding up leads to
\begin{eqnarray}
S^{\mathrm{bl}}_{L}(2n) 
& = \Delta_L H^+_L
 + \Phi^+_L(L/2-n) + \Phi^+_L(n)
 +  \frac{1}{2} G_L(2n) + \frac{1}{2} \Gamma_L .
\end{eqnarray}
%
%
A similar computation for odd distance $r=2n-1$ yields
\begin{eqnarray}
S^{\mathrm{bl}}_{L}(2n-1) 
& =  \Delta_L H^-_L 
+ \Phi^-_L(L/2-n) + \Phi^-_L(n-1) + \frac{1}{2} G_L(2n-1) \nonumber \\ 
& - \frac{\Delta_L}{L} \left(\tilde{H}_L(n) Q^+_{n}
+ H_L(n) Q^-_{n} \right) .
\end{eqnarray}
\normalsize
In the thermodynamic limit for $z\neq z_c$ with $r$ fixed each term
vanishes individually which yields 
$S^{\mathrm{bl}}(r) = 0 $, thus completing the
proof of Theorem \ref{Theo:Ossf}. On the other hand, for
hydrodynamic scaling one obtains \eref{Sblu}, thus completing the
proof of Theorem \ref{Theo:Cssf}.

\paragraph{\underline{Proof of Theorem \ref{Theo:Pnf}:}}

The result \eref{Chc} follows directly from \eref{Bhc}.
%
From \eref{SLrrefl} one gets
\begin{eqnarray}
C^{\mathrm{refl}}_L
& = - \frac{2 j_L}{p L} \sum_{n=1}^{L/4} 
\left[\tilde{H}_L(n) Q^+_{n} + H_L(n)  Q^-_{n}\right].
\end{eqnarray}
Taking the thermodynamic limit yields \eref{Crefl}.
The leading contribution from the boundary layer part is the sum
$\sum_{r=1}^{L/2} G_L(r)$ which vanishes for for $z=\neq z_c$ in the
thermodynamic limit and for $z=z_c$ is proportional to $L$ since each
term in the sum is of order 1. Hence the limit $\lim_{L\to\infty} C^{\mathrm{refl}}_L$ diverges for $z=z_c$ which proves \eref{Cbl}.

From the scaling form \eref{Sblu} one calculates
\begin{equation}
\lim_{L\to\infty} C^{\mathrm{bl}}_L/L = 2 \int_0^{1/2} 
S^{\mathrm{bl}}(u) \rmd u
\end{equation}
which yields \eref{sCbl}. \qed


\appendix

\section{Auxiliary constants and functions}
\label{App:lim}

We collect exact finite-size expressions, the asymptotic behaviour for
large $L$, and the thermodynamic limits $L\to\infty$ of 
constants and functions that are used in the proofs throughout this paper. 
The symbol $\epsilon(L)$ denotes unspecified corrections exponentially 
small in $L$ that may differ from formula to formula. 

For $z\neq z_c$ we define the quantity
\begin{equation}
\label{loclengthdef1}
\xi_s(p,z) := \left(\ln{\frac{z_c(1+z)}{z(1+z_c)}}\right)^{-1}
\end{equation}
that appears in the normalization ratio \eref{YMN}.
For $z>z_c$ one has $\xi_s<0$ which needs to be borne in mind when taking 
thermodynamic limits. For finite size
we use the parameter $\xi_s$ in the expressions below, while for
large $L$ and for the thermodynamic limit we express all results
in terms of the localization length $\xi = |\xi_s|$.

\subsection{The effective length $L_{\mathrm{eff}}$}

For $p \neq 1$ one reads off from the definition \eref{Leffdef}
\begin{equation}
\frac{1}{L_{\mathrm{eff}}} 
= \frac{1}{L} - \frac{2}{1-p} \frac{1}{L^2} + O(L^{-3}).
\end{equation}
For the limiting case $p=1$ we note that $\lim_{p\to 1} (1-p) L_{\mathrm{eff}}
=2$ which implies that taking the thermodynamic limit $L\to\infty$ and the 
limit $p\to1$ cannot be interchanged whenever $L_{\mathrm{eff}}$
appears in finite-size expressions.

\subsection{The functions $H_L(n)$ and $\tilde{H}_L(n)$}
\label{App:H}

\paragraph{Exact finite size expression:}

According to the definition (\ref{loclengthdef}) one has
\begin{equation}
\rme^{-1/\xi_s} = \frac{z(1+z_c)}{z_c(1+z)}
\end{equation}
where
\begin{equation}
\rme^{-1/\xi_s} = \left\{ 
\begin{array}{ll}
< 1 & z \neq z_c \\
1 & z = z_c \\
> 1 & z \neq z_c .
\end{array}
\right.
\end{equation}

From the definitions \eref{HLdef} one gets
\begin{equation}
H_L(n) = p(1-p) \left\{ 
\begin{array}{ll}
\displaystyle 
\frac{\rme^{-n/\xi_s} - \frac{z}{z_c} \rme^{-L/(2\xi_s)}}
{1 - \frac{z}{z_c} \rme^{-L/(2\xi_s)}} & \qquad z \neq z_c \\[4mm]
\displaystyle \left(1 - \frac{2n}{L_{\mathrm{eff}}}\right) & 
\qquad z = z_c 
\end{array}
\right.
\label{HLn}
\end{equation}
\begin{equation}
\tilde{H}_L(n) = p(1-p) \left\{ 
\begin{array}{ll}
\displaystyle 
\frac{1 - \frac{1+z_c}{1+z} \rme^{n/\xi_s}}
{1 - \frac{z_c}{z} \rme^{L/(2\xi_s)}} & \qquad z \neq z_c \\[4mm]
\displaystyle \frac{ 2(n+z_c))}{L_{\mathrm{eff}}} & 
\qquad z = z_c 
\end{array}
\right.
\label{tHLn}
\end{equation}
and for the lattice sums 
\begin{eqnarray}
H^{\pm}_L := \frac{2}{L} \sum_{k=1}^{\frac{L}{2}} Q^\pm_k H_L(k) 
\label{HLpmdef}
\end{eqnarray}
one finds
\begin{eqnarray}
H^{+}_L
& = p(1-p) \left\{ 
\begin{array}{ll}
\displaystyle \frac{\frac{2}{L}\frac{1-\rme^{-L/(2\xi_s)}}{\rme^{2/\xi_s}-1} 
- \frac{z}{z_c} \rme^{-L/(2\xi_s)}}
{1 - \frac{z}{z_c} \rme^{-L/(2\xi_s)}} 
& \qquad z \neq z_c \\[4mm]
\displaystyle \frac{1}{4} 
\left(1 + \frac{2(2p-1)}{(1-p)L_{\mathrm{eff}}}\right) & \qquad z = z_c ,
\end{array}
\right.
\label{HLsumeven}
\end{eqnarray}

\begin{eqnarray}
H^{-}_L
& = p(1-p) \left\{ 
\begin{array}{ll}
\displaystyle \frac{\frac{2}{L}\frac{1-\rme^{-L/(2\xi_s)}}{\rme^{1/\xi_s}-\rme^{-1/\xi_s}}
- \frac{z}{z_c} \rme^{-L/(2\xi_s)}}
{1 - \frac{z}{z_c} \rme^{-L/(2\xi_s)}} 
& \qquad z \neq z_c \\[4mm]
\displaystyle \frac{1}{4} 
\left(1 + \frac{2}{(1-p)L_{\mathrm{eff}}}\right) & \qquad z = z_c .
\end{array}
\right.
\label{HLsumodd}
\end{eqnarray}
We remark that according to \ref{rhoLn} these lattice sums are related to the sublattice 
densities
\begin{eqnarray}
\frac{2}{L} \exval{N^+}_L
& = \frac{1}{p} H^{-}_L
\label{rhoevenL} \\
\frac{2}{L} \exval{N^-}_L
& = j_L + \frac{1}{p} H^{+}_L.
\label{rhooddL} 
\end{eqnarray}

\paragraph{Asymptotic behaviour for large $L$:}

For fixed $n$ one obtains
\begin{equation}
H_L(n) = p(1-p) \left\{ 
\begin{array}{ll}
\displaystyle
\rme^{-n/\xi} + \epsilon(L)
& \qquad z < z_c \\[4mm]
\displaystyle
1 - \frac{2n}{L} + O(L^{-2}) 
& \qquad z = z_c \\[4mm]
\displaystyle
1 + \epsilon(L)
& \qquad z > z_c .
\end{array}
\right.
\label{HLasymp1}
\end{equation}

\begin{equation}
\tilde{H}_L(n) = p(1-p) \left\{ 
\begin{array}{ll}
\displaystyle
\epsilon(L)
& z < z_c \\[4mm]
\displaystyle
\frac{2n+2z_c}{L} + O(L^{-2}) & z = z_c \\[4mm]
\displaystyle
1 - \frac{1+z_c}{1+z} \rme^{-n/\xi} + \epsilon(L)
& z > z_c .
\end{array}
\right.
\label{HLasymp2}
\end{equation}

The lattice sums behave asymptotically as
\begin{eqnarray}
H^{+}_L
& = p(1-p) \left\{ 
\begin{array}{ll}
\displaystyle \frac{2\rme^{-2/\xi}}{1-\rme^{-2/\xi}} \frac{1}{L} + \epsilon(L)
& \qquad z < z_c \\[4mm]
\displaystyle \frac{1}{4} 
- \frac{1-2p}{2(1-p)} \frac{1}{L} 
+ O(L^{-2})
& \qquad z = z_c  \\[4mm]
\displaystyle 1 - \frac{2}{1-\rme^{-2/\xi}} \frac{z_c}{z} \frac{1}{L}
+ \epsilon(L)
& \qquad z > z_c ,
\end{array}
\right.
\label{HLsumevenasymp}
\end{eqnarray}

\begin{eqnarray}
H^{-}_L
& = p(1-p) \left\{ 
\begin{array}{ll}
\displaystyle \frac{2 \rme^{-1/\xi}}{1-\rme^{-2/\xi}} \frac{1}{L} + \epsilon(L)
& \qquad z < z_c \\[4mm]
\displaystyle \frac{1}{4} + \frac{1}{2(1-p)} \frac{1}{L} + O(L^{-2})
& \qquad z = z_c \\[4mm]
\displaystyle 1 - \frac{2 \rme^{-1/\xi}}{1-\rme^{-2/\xi}}
\frac{z_c}{z} \frac{1}{L}
& \qquad z > z_c .
\end{array}
\right.
\label{HLsumoddasymp}
\end{eqnarray}

\paragraph{Thermodynamic limit for fixed $n\in\Z$:}

In the thermodynamic limit one has for $z\neq z_c$ and $n\in\N_0$
\begin{equation}
H(n) := \lim_{L\to\infty} H_L(n) = p(1-p) \left\{ 
\begin{array}{ll}
\rme^{-n/\xi} & z < z_c \\[4mm]
1 & z \geq z_c .
\end{array}
\right.
\label{H}
\end{equation}
\begin{equation}
\tilde{H}(n) := \lim_{L\to\infty} \tilde{H}_L(n) = p \left\{ 
\begin{array}{ll}
\displaystyle 0 & z \leq z_c \\[4mm]
\displaystyle 1-p - \frac{\rme^{-n/\xi}}{1+z}  & z > z_c .
\end{array}
\right.
\label{tH}
\end{equation}
At the critical point one gets for $u\in[0,1/2]$
\begin{equation}
H_c(u) := \lim_{L\to\infty} H_L(\floor{uL}) \Bigr|_{z = z_c} = p(1-p)(1 - 2u)  .
\label{Hc}
\end{equation}
\begin{equation}
\tilde{H}_c(u) := \lim_{L\to\infty} H_L(L/2+1-\floor{uL}) \Bigr|_{z = z_c} 
= p(1-p) 2u
\label{tHc}
\end{equation}

For the lattice sums one gets
\begin{eqnarray}
H^{\pm} := \lim_{L\to\infty} H^{\pm}_L
& = p(1-p) \left\{ 
\begin{array}{ll}
\displaystyle 0
& \qquad z < z_c \\[4mm]
\displaystyle \frac{1}{4} 
& \qquad z = z_c  \\[4mm]
\displaystyle 1 
& \qquad z > z_c .
\end{array}
\right.
\label{Hsum}
\end{eqnarray}

\subsection{The constants $\Gamma_L$  and 
$\Delta_L$}

Using \eref{jL} one obtains from the definitions \eref{GammaLdef} and 
\eref{DeltaLdef} the exact expressions
\begin{eqnarray}
\Gamma_L & = \cases{ - \frac{\left(\frac{z}{1+z} - 
\frac{z_c}{1+z_c} \right)^2 \frac{z}{z_c} \rme^{-L/(2\xi_s)}}
{\left(1 - \frac{z}{z_c} \rme^{-L/(2\xi_s)}\right)^2} 
& $z\neq z_c$ \\
 - \frac{4 p^2}{L_{\mathrm{eff}^2}} . & $z = z_c$ 
}
\label{GammaL} \\
\Delta_L & = \cases{
\frac{1 - \rme^{-1/\xi_s}}{1 - \left(\frac{z}{z_c}\right) \rme^{-L/(2\xi_s)}} 
& $z \neq z_c$ \\
\frac{2}{L_{\mathrm{eff}}} & $z = z_c$.
} 
\label{DeltaL} 
\end{eqnarray}
It follows that
\begin{eqnarray}
\Delta & := \lim_{L\to\infty} \Delta_L = \cases{
1 - \frac{1}{p} \frac{z}{(1+z)}
& $z < z_c$ \\
0 & $z \geq z_c$} 
\label{Delta} \\
\Gamma & := \lim_{L\to\infty} \Gamma_L = 0 .
\label{Gamma} 
\end{eqnarray}

\subsection{The functions $\Psi_L(n)$ and $\tilde{\Psi}_L(n)$}

From the definition \eref{PsiLdef} one finds
with \eref{Leffdef} and \eref{DeltaL} the exact expressions
\begin{eqnarray}
\Psi_L(m) & = - p(1-p) \Delta_L \rme^{-m/\xi_s} 
\label{PsiL} \\
\tilde{\Psi}_L(m) & = - p(1-p) \Delta_L \rme^{-L/(2\xi_s)} \rme^{(m-1)/\xi_s} 
\label{tPsiL}
\end{eqnarray}
and, for $m \in \N$ fixed, the limits
\begin{eqnarray}
\Psi(m) & := \lim_{L\to\infty} \Psi_L(m) = \left\{ 
\begin{array}{ll}
- p \left(1-\rme^{-1/\xi_s}\right) \frac{\rme^{-m/\xi}}{1+z_c} 
& z < z_c \\[4mm]
0 & z \geq z_c 
\end{array}
\right. 
\label{Psi} \\
\tilde{\Psi}(m) & := \lim_{L\to\infty} \tilde{\Psi}_L(m) = \left\{ 
\begin{array}{ll}
0
& z \leq z_c \\[4mm]
- p \left(\rme^{1/\xi}-1\right) \frac{\rme^{-m/\xi}}{1+z}  
& z > z_c .
\end{array}
\right. 
\label{tPsi} 
\end{eqnarray}
with $\Delta$ obtained in \eref{Delta}.

Away from the critical point the finite-size corrections to these asymptotic values
are exponentially small in $L$. For the critical point we note that 
\begin{eqnarray}
\Psi(m) & = - \tilde{\Psi}(m) = -2 p(1-p) \frac{1}{L} + O(L^{-2}).
\end{eqnarray}

Evaluating the sums $\Phi^\pm(n)$ defined by \eref{PhiLndef} yields
\begin{eqnarray}
\Phi^+_L(m) & =  \frac{1}{L} \frac{H_L(2\floor{m/2}) - H_L(2)}{1 + \rme^{-1/\xi_s}} \\
\Phi^-_L(m) & = \frac{1}{L} \frac{H_L(2\floor{(m+1)/2}-1) - H_L(1)}{1 + \rme^{-1/\xi_s}} \\
\end{eqnarray}
and $\lim_{L\to\infty} \Phi^\pm_L(m) = 0$ with corrections of order $1/L$
to the asymptotic result.

\subsection{The function $F_L(m,n)$ and $\tilde{F}(m,n)$}

From the definition \eref{FLmndef} and the exact expression \eref{HLn} one 
gets
\begin{eqnarray}
F_L(m,n) & = \left\{ 
\begin{array}{ll}
(1-p)^2 \frac{\left(\rme^{-m/\xi_s} - \frac{z}{z_c} \rme^{-L/(2\xi_s)}\right)
\left(1 - \rme^{-n/\xi_s}\right)}
{\left(1 - \frac{z}{z_c} \rme^{-L/(2\xi_s)}\right)^2} 
& z \neq z_c \\[4mm]
(1-p)^2 \left(1 - \frac{2m}{L_{\mathrm{eff}}}\right) \frac{2n}{L_{\mathrm{eff}}} & z = z_c 
\end{array}
\right. 
\label{FLmn} \\
\tilde{F}_L(m,n) & = \left\{ 
\begin{array}{ll}
(1-p)^2 \frac{\left(1 - \frac{1+z_c}{1+z} \rme^{m/\xi_s}\right)
\left(\frac{1+z_c}{1+z} \rme^{n/\xi_s} - \frac{z_c}{z} \rme^{L/(2\xi_s)}\right)}
{\left(1 - \frac{z_c}{z}\rme^{L/(2\xi_s)}\right)^2} 
& z \neq z_c \\[4mm]
(1-p)^2 \left(1 - \frac{2n + 2 z_c}{L_{\mathrm{eff}}}\right) 
\frac{2m + 2z_c}{L_{\mathrm{eff}}} & z = z_c 
\end{array}
\right. 
\label{tFLmn}
\end{eqnarray}
and, for $m,n \in \N$ fixed, the limits
\begin{eqnarray}
F(m,n) & := \lim_{L\to\infty} F_L(m,n)  = \left\{ 
\begin{array}{ll}
(1-p)^2 \rme^{-m/\xi}
\left(1 - \rme^{-n/\xi}\right)
& z < z_c \\[4mm]
0 & z \geq z_c 
\end{array}
\right.
\label{Fmn} \\
\tilde{F}(m,n) & := \lim_{L\to\infty} \tilde{F}_L(m,n)  = \left\{ 
\begin{array}{ll}
0 & z \leq z_c \\[4mm]
\frac{\rme^{-n/\xi}}{1+z}  \left(1-p - \frac{\rme^{-m/\xi}}{1+z} 
\right) & z > z_c .
\end{array}
\right.
\label{tFmn} 
\end{eqnarray}
Away from the critical point the finite-size corrections to these asymptotic values
are exponentially small in $L$. 

For $z\neq z_c$ one has
\begin{eqnarray}
G_L(r) 
& = \frac{2(1-p)^2}{L\left(1 - \frac{z}{z_c} \rme^{-L/(2\xi_s)}\right)^2} 
\rme^{-1/\xi_s} \left[
\frac{\rme^{-r/\xi_s} - \rme^{-L/(2\xi_s)}}{1-\rme^{-1/\xi_s}}
\right. \nonumber \\
& \left. + \rme^{-1/\xi_s}
\frac{\rme^{-r/\xi_s} - \rme^{-(L-r)/\xi_s}}{1-\rme^{-2/\xi_s}}
 - \frac{z}{z_c} 
\frac{\rme^{-L/(2\xi_s)} - \rme^{-(L-r)/\xi_s}}{1-\rme^{-1/\xi_s}}
\right] \nonumber \\
& - \left(1- \frac{2r}{L}\right)\frac{(1-p)^2}{\left(1 - \frac{z}{z_c} \rme^{-L/(2\xi_s)}\right)^2} \frac{z}{z_c} \rme^{-L/(2\xi_s)} 
\end{eqnarray}

For $z= z_c$ one has
\begin{eqnarray}
G_L(r) 
& = \frac{1}{6} (1-p)^2 \left(1 - \frac{2r}{L}\right) 
\left(1-\frac{2r+2z_c}{L_{\mathrm{eff}}}\right) 
\left(1 - \frac{2r-4z_c-2}{L_{\mathrm{eff}}}\right)
\end{eqnarray}

Asymptotically this yields
\small
\begin{eqnarray}
G_L(r) 
& = \cases{\frac{2}{L}  \frac{z}{(1+z_c)(z_c - z)}
\left(1 + \frac{z(1+z_c)}{z+z_c+2 z z_c}
\right) \rme^{-r/\xi} +\epsilon_L & $z<z_c$ \\
\frac{1}{6} (1-p)^2 \left(1 - \frac{2r}{L}\right)^3 + O(1/L) & $z=z_c$ \\
\frac{2}{L} \frac{- z_c}{(1+z_c)(z - z_c)} \left[1 - 
\frac{z_c^2(1+z)}{z(z+z_c+2 z z_c)} \right] \rme^{-r/\xi}  +\epsilon_L
& $z>z_c$ .
} \nonumber \\
\end{eqnarray}
\normalsize
Thus for the limit
\begin{equation}
G_\infty(r) := \lim_{L\to\infty} G_L(r), \qquad r \in \N
\end{equation}
one gets for fixed $r$
\begin{eqnarray}
G_\infty(r) 
& = \cases{0 & $z<z_c$ \\
\frac{1}{6} (1-p)^2 & $z=z_c$ \\
0 & $z>z_c$
} 
\end{eqnarray}
while for the limit
\begin{equation}
G(u) := \lim_{L\to\infty} G_L(\floor{uL}), \qquad u \in [0,1/2]
\end{equation}
one finds
\begin{eqnarray}
G(u) 
& = \cases{0 & $z<z_c$ \\
\frac{1}{6} (1-p)^2 (1-2u)^3 & $z=z_c$ \\
0 & $z>z_c$.
} 
\end{eqnarray}

\section{On the matrix product ansatz for the dsTASEP}
\label{App:MPA}

We briefly explain how the MPA of \cite{Hinr97} is related to the SMPM
(\ref{MPA}). We consider $L/2$ even and recall the notation 
\begin{equation}
\sigma_{k} := \eta_{k} + 2 \eta_{L+1-k}, \quad 1 \leq k \leq \frac{L}{2}
\end{equation}
of \cite{Hinr97} that represents the occupation pair $(\eta_{L+1-k},\eta_{k})$ as
\begin{eqnarray}
(0,0) \mapsto 0, \qquad
(0,1) \mapsto 1, \qquad
(1,0) \mapsto 2, \qquad
(1,1) \mapsto 3.
\end{eqnarray}
The state $\eta$ of the dsTASEP can thus be expressed in terms of the state 
variable $\sigma = (\sigma_1,\dots ,\sigma_{L/2})$. One has
\begin{equation}
\delta_{\sigma_{k},\sigma} = \frac{1}{4} \sum_{j=0}^{3} 
\rme^{i \frac{\pi}{2}j(\sigma_{k}-\sigma)}, \quad 1 \leq k \leq \frac{L}{2}
\end{equation}
and therefore
\begin{eqnarray}
\eta_{k} & = \left(\delta_{\sigma_{k},1} + \delta_{\sigma_{k},3}\right)
\Theta^{(1)}_{L,k} + \left(\delta_{\sigma_{L+1-k},2} + \delta_{\sigma_{L+1-k},3}\right) \Theta^{(2)}_{L,k} \\
\bar{\eta}_{k} & = \left(\delta_{\sigma_{k},0} + \delta_{\sigma_{k},2}\right)
\Theta^{(1)}_{L,k} + \left(\delta_{\sigma_{L+1-k},0} + \delta_{\sigma_{L+1-k},1}\right) \Theta^{(2)}_{L,k} .
\end{eqnarray}

According to \cite{Hinr97} the MPA is given in terms of vectors $\langle W|$, 
$| V \rangle$, matrices $A_\sigma$, $B_\sigma$, and a normalization factor 
$Y_L$ by 
\begin{equation}
P(\sigma) = \frac{1}{Y_L} 
\langle W | A_{\sigma_1} B_{\sigma_2} A_{\sigma_3} B_{\sigma_4}
\dots A_{\sigma_{L/2-1}} B_{\sigma_{L/2}} | V \rangle 
\end{equation}
where the quantum meachnical bra-ket convention for scalar products is
used. In terms of the occupation variables $\eta_k$ this reads
\begin{eqnarray}
P(\eta) & = \frac{1}{Y_L} \langle W | 
(\bar{\eta}_1\bar{\eta}_L A_0 + \bar{\eta}_1 \eta_L A_2 
+ \eta_1\bar{\eta}_L A_1 + \eta_1\eta_L A_3) \nonumber \\
& \times
(\bar{\eta}_{2}\bar{\eta}_{L-1} B_0 + \bar{\eta}_{2} \eta_{L-1} B_2 
+ \eta_{2}\bar{\eta}_{L-1} B_1 + \eta_{2}\eta_{L-1} B_3)\nonumber \\
& \times
\dots \nonumber \\
& \times
(\bar{\eta}_{\frac{L}{2}-1}\bar{\eta}_{\frac{L}{2}+2} A_0 + \bar{\eta}_{\frac{L}{2}-1} \eta_{\frac{L}{2}+2} A_2 
+ \eta_{\frac{L}{2}-1}\bar{\eta}_{\frac{L}{2}+2} A_1 + \eta_{\frac{L}{2}}\eta_{\frac{L}{2}+2} A_3)\nonumber \\
& \times
(\bar{\eta}_{\frac{L}{2}}\bar{\eta}_{\frac{L}{2}+1} B_0 + \bar{\eta}_{\frac{L}{2}} \eta_{\frac{L}{2}+1} B_2 
+ \eta_{\frac{L}{2}}\bar{\eta}_{\frac{L}{2}+1} B_1 + \eta_{\frac{L}{2}}\eta_{\frac{L}{2}+1} B_3)
| V \rangle .
\end{eqnarray}
The vectors are
\begin{equation}
\langle W | = (1,1), \qquad 
| V \rangle = \left(\begin{array}{c} 1 \\ 1 \end{array}\right)
\end{equation}
and one has $B_1=B_3=0$, $B_0 = A_0$, and $B_2 = A_1 + A_2$. 
Multiplying the matrices $A_\sigma$ of \cite{Hinr97} by a factor of $p$,
and replacing the scalar product by the trace involving the Kronecker product 
$D = (| V \rangle \otimes \langle W |)/2$ one arrives at (\ref{MPA}).

\end{document}